\documentclass[a4paper,fleqn]{cas-dc}

\usepackage[authoryear,longnamesfirst]{natbib}
\usepackage{oplotsymbl}
\usepackage{tikz}

\def\tsc#1{\csdef{#1}{\textsc{\lowercase{#1}}\xspace}}
\tsc{WGM}
\tsc{QE}
\tsc{EP}
\tsc{PMS}
\tsc{BEC}
\tsc{DE}

\newcommand{\reva}[1]{\textcolor{black}{#1}}
\newcommand{\revb}[1]{\textcolor{black}{#1}}

\begin{document}
\let\WriteBookmarks\relax
\def\floatpagepagefraction{1}
\def\textpagefraction{.001}

\shorttitle{Laminar planar jets of elastoviscoplastic fluids}

\shortauthors{G Soligo, ME Rosti}

\title [mode = title]{Laminar planar jets of elastoviscoplastic fluids}                      



%
\author{Giovanni Soligo}[orcid=0000-0002-0203-6934]
\cormark[1]


\ead{soligo.giovanni@oist.jp}



\affiliation{organization={Okinawa Institute of Science and Technology Graduate University},
    addressline={Tancha 1919-1}, 
    city={Onna},
    postcode={904-0495}, 
    state={Okinawa},
    country={Japan}}

\author{Marco Edoardo Rosti}[orcid=0000-0002-9004-2292]
\cormark[1]
\ead{marco.rosti@oist.jp}

\cortext[cor1]{Corresponding author}



\begin{abstract}
We perform numerical simulations of planar jets of elastoviscoplastic (EVP) fluid (\citet{saramito2007new} model) at low Reynolds number. Three different configurations are considered: $(i)$ EVP jet in EVP ambient fluid, $(ii)$ EVP jet in Newtonian ambient fluid (miscible), and $(iii)$ EVP jet in Newtonian ambient fluid (immiscible). We investigate the effect of the Bingham number, \textit{i.e.} of the dimensionless yield stress of the EVP fluid, on the jet dynamics, and find a good agreement with the scaling for laminar, Newtonian jets for the centerline velocity $u_c/u_0\propto (x/h)^{-1/3}$ and for the jet thickness $\delta_{mom}/h\propto (x/h)^{2/3}$ at small Bingham number. This is lost once substantial regions of the fluid become unyielded, where we find that the spreading rate of the jet and the decay rate of the centerline velocity increase with the Bingham number, due to the regions of unyielded fluid inducing a blockage effect on the jet. The most striking difference among the three configurations we considered is the extent and position of the regions of unyielded fluid: large portions of ambient and jet fluids (in particular, away from the inlet) are unyielded for the EVP jet in EVP ambient fluid, whereas for the other two configurations, the regions of unyielded fluid are limited to the jet, as expected. We derive a power law scaling for the centerline yield variable and confirm it with the results from our numerical simulations. The yield variable determines the transition from the viscoelastic Oldroyd-B fluid (yielded) to the viscoelastic Kelvin-Voigt material (unyielded).
\end{abstract}


\begin{highlights}
\item The formation of unyielded fluid regions increases the spreading rate of the jet and the decay rate of the centerline velocity.
\item Laminar power law scalings for Newtonian planar jets apply to elastoviscoplastic jets, for low Bingham numbers.
\item The yield variable at the centerline scales as the local reference stress, computed using the jet thickness and the centerline velocity, used as reference length and velocity scales.
\end{highlights}

\begin{keywords}
planar jet \sep elastoviscoplastic fluid \sep Saramito model 
\end{keywords}

\maketitle

\section{Introduction}


Yield stress fluids are a category of non-Newtonian fluids characterized by the presence of a yield stress: these fluids deform as a solid when the applied stress is lower than the yield stress, and flow as a liquid when the applied stress is larger than the yield stress.
Examples of yield stress fluids range from household products, such as ketchup, mayonnaise and toothpaste, to fluids relevant in industrial applications, such as cements, mortars, foams and crude oils, and in environmental flows, such as mud, magma and basaltic lava flows \citep{balmforth2014yielding,coussot2014yield,frigaard2019simple,malkin2017modern,bird1983rheology,nguyen1992measuring}. 
Yield stress fluids find use also in bio-medical applications: \citet{schramm2004jet} used a polyacrylamide gel (yield stress fluid) as a model for human skin in their studies on high-speed drug injection into human skin.

The peculiar behaviour of these fluids originates from their dual nature of fluid and solid: toothpaste can be squeezed out of the tube and laid on the toothbrush (fluid behavior) and keeps its shape once on the toothbrush (solid behavior).
Understanding the dynamics of yield stress fluids is a challenging task; \citet{balmforth2014yielding} reviewed recent advances in the modeling of yield stress fluids, whereas \citet{frigaard2019simple} provided a critical comparison of newer and original models of yield stress fluids.
Experimental efforts in characterizing the properties of yield stress fluids via rheometric measurements have been instead reviewed by \citet{coussot2014yield, nguyen1992measuring}. An ongoing source of debate concerns the existence of a sharp transition at a well-defined yield stress \citep{malkin2017modern}.

The effect of the fluid plasticity in wall-bounded flows has been unveiled in previous numerical works with an elastoviscoplastic fluid: \citet{rosti2018turbulent} reported an increase in the friction factor in laminar conditions for increasing values of the Bingham number, representative of the dimensionless yield stress of the fluid. Conversely, it was observed that in turbulent wall-bounded flows the friction factor decreases with increasing Bingham numbers \citep{rosti2018turbulent,izbassarov2021effect}; using high-order dynamic mode decomposition to study the evolution of flow structures, \citet{le2020coherent} demonstrated that the lower friction factor, \textit{i.e.} the lesser drag, is linked to the re-laminarisation of the flow. Larger structures moving at a lower frequency cause a regularisation of the flow, and thus the observed drag reduction. Experiments from \citet{mitishita2021fully} confirmed these findings: the addition of Carbopol resulted in an increase in the stream-wise component of the Reynolds stresses and in a reduction of the wall-normal component. The combined effect of plasticity and fluid elasticity was investigated by \citet{izbassarov2021effect}; it was demonstrated that, once the flow re-laminarises, an increase in plasticity leads to drag increase for viscoplastic fluids, whereas for elastoviscoplastic fluids drag reduction is still observed.

Understanding the flowing dynamics of yield stress fluids is of utmost importance in the oil and gas industry: the final stage of an oil and gas extraction well is the plug and abandonment step. At the end of the well lifespan, it becomes economically unattractive to continue the extraction of gas and oil, and the well has to be sealed to prevent water source contamination and leakages of hydrocarbons  \citep{Nelson2006,khalifeh2020introduction,akbari2021fluid}. The well is sealed using a concrete plug and, to make sure the concrete can set properly and and the plug is structurally solid, slurries and drilling muds have to be washed away \citep{hassanzadeh2023neutrally}. These fluids behave as yield stress fluids, and are either washed away using water jets or displaced via injection of concrete \citep{akbari2021fluid,eslami2022experimental,kazemi2024buoyant}.
Prototypical experiments for the plug and abandonment process investigate the displacement of Newtonian fluid by a heavier yield stress fluid, and include dam-breaking \citep{kazemi2024buoyant} and annular injection \citep{akbari2022immersed, akbari2022breakup, akbari2021fluid, eslami2022experimental} configurations. 
Some of these configurations are akin to jet flows: \citet{hassanzadeh2023neutrally} use water jets (Newtonian) to displace a yield stress fluid, while others studied the injection of a yield stress fluid to displace a Newtonian fluid \citep{akbari2022immersed, akbari2022breakup, akbari2021fluid, eslami2022experimental}; however, in these latter cases the relative density between the two fluids is the driving force and the focus of the study is on the breakage of the column of yield stress fluid and on its deposition.
Yield stress fluids can also be found in 3D printing applications, as a support for the printed object \citep{wu2022suspension} or as printing material \citep{geffrault2023printing, mackay2018importance, hua2021numerical, karyappa2019immersion, lawson2021recent}. In this latter case, the problem configuration is very similar to the one considered in this work: the injection of a yield stress fluid.


Despite the importance of this type of flow configuration, \textit{i.e.} the injection of a yield stress fluid in a generic ambient fluid, there are relatively few works aimed at understanding the basic flow dynamics of a jet of yield stress fluid, as most of the archival literature mainly focuses on the dynamics of deposition of a yield stress fluid on a solid surface (buckling, coiling, mounding) \citep{balmforth2014yielding, akbari2022breakup}.
\citet{maleki2018submerged} analysed the efficiency of waste-water jet mixing (sludge re-circulation) via numerical simulations as a function of the flowing regime, yield stress of the fluid, and digester design. It was found that for low-yield stress fluids an increase in the Reynolds number causes a strong reduction in the size of regions of stagnant (unmixed) fluid, thus improving the efficiency of the digester. \citet{maleki2018submerged} also suggested that designing the digester by following the shape of the yielded fluid massively reduces the regions characterized by stagnant fluid.

In this work we provide an extensive characterization of a low-Reynolds number jet of elastoviscoplastic fluid, defined by the Saramito fluid model \citep{saramito2007new}.
We investigate the effect of the Bingham number on the flow dynamics for three different configurations: $(i)$ EVP jet fluid in EVP ambient fluid, $(ii)$ EVP jet fluid in Newtonian ambient fluid (with the EVP and Newtonian fluids being miscible), and $(iii)$ EVP jet fluid in Newtonian ambient fluid (with the EVP and Newtonian fluids being immiscible). 
We present the classical flow jet statistics, and characterize the non-Newtonian extra-stresses and unyielded fluid regions. 
We observe, in all cases, that an increase in the Bingham number results in an increased spreading rate of the jet and decay rate of the centerline velocity, and in a shortening of the potential core of the jet. The latter result complements \citet{kumar1984laminar} experimental measurements of the laminar length of non-Newtonian jets, where an increase of the laminar length with the Reynolds number is reported for $Re<200$. 

The paper is structured as follows. First, in section~\ref{sec: nummet}, we present the method that has been used for all numerical simulations presented in this work. The computational setup is introduced together with all the relevant parameters of the problem and the list of simulations in section~\ref{sec: setup}. Results are reported in section~\ref{sec: results}, starting from the classic jet statistics, \textit{i.e.} jet thickness and centerline velocity, and following with additional details on the non-Newtonian extra-stress and on the yielded/unyielded regions of the flow. Finally, section~\ref{sec: concl} summarizes the main findings. 

%
%

\section{Mathematical model and numerical method}
\label{sec: nummet}

We perform direct numerical simulations of a non-Newtonian jet at low Reynolds number. The non-Newtonian fluid is defined by the elastoviscoplastic (EVP) Saramito model \citep{saramito2007new}; this fluid model is characterized by fluid elasticity and plastic behaviour. The fluid behaves as a viscoelastic Kelvin-Voigt before yielding (i.e. when the magnitude of the deviatoric stress is lower than the yield stress), and as a viscoelastic Oldroyd-B-like fluid once yielded. The value of the yield stress determines the transition between the two behaviours. 
We examine three different non-Newtonian configurations: one-fluid (EVP jet in EVP ambient fluid), two-fluid (EVP jet in Newtonian ambient fluid) miscible (EVP and Newtonian fluids can mix) and two-fluid immiscible (EVP and Newtonian fluids keep separate). An additional Newtonian case, \textit{i.e.} Newtonian fluid in Newtonian ambient fluid, has been also simulated for the sake of comparison.

The dynamic of the system is defined by the incompressibility constraint and the conservation of momentum, defined as
\begin{equation}
\nabla \cdot \boldsymbol{u} =0,
\label{eq:cont}
\end{equation}
\begin{equation}
\rho \left(\frac{\partial \mathbf{u}}{\partial t} + \mathbf{u}\cdot\nabla{\mathbf{u}}\right) = -\nabla p + \nabla \cdot \left[ \eta_s \left( \nabla\mathbf{u}+\nabla\mathbf{u}^T \right) + \alpha\boldsymbol{\tau} \right].
\label{eq:ns}
\end{equation}
In the above equation, $\rho$ is the fluid density, $\eta_s$ the solvent viscosity, $\mathbf{u}$ the local fluid velocity, $p$ the pressure, and $\boldsymbol{\tau}$ the non-Newtonian extra-stress tensor (which is absent in the Newtonian case).
The variable $\alpha$ defines the local volume fraction of the EVP fluid: for the one-fluid case $\alpha=1$ everywhere, for the two-fluid miscible case $\alpha$ is equal to the local concentration $c$, and for the two-fluid immiscible case $\alpha$ is equal to the volume of fluid variable $\phi$.
The non-Newtonian extra-stress tensor $\tau$ is advanced in time by solving the following additional transport equation:
\begin{equation}
\lambda \overset{\nabla}{\boldsymbol{\tau}}+\max\left(0, 1-\alpha\tau_y/|\tau_d| \right)\boldsymbol{\tau}=\eta_p\left( \nabla\mathbf{u} +\nabla\mathbf{u}^T \right),
\label{eq:nntau}
\end{equation}
where $\lambda$ is the relaxation time scale, $\tau_y$ the yield stress, $|\tau_d|$ the magnitude of the deviatoric stress tensor, and $\eta_p$ the polymer viscosity. \revb{The yield stress is weighted by the local volume fraction, similarly to what done in equation~\ref{eq:ns} for the polymeric extra-stress contribution.} The magnitude of the deviatoric stress tensor is defined as $|\tau_d|=\sqrt{\boldsymbol{\tau}_d:\boldsymbol{\tau}_d}$, with $\boldsymbol{\tau}_d=\boldsymbol{\tau}-\text{tr}(\boldsymbol{\tau})\boldsymbol{I}/3$, being $\boldsymbol{I}$ the identity tensor.
The upper-convected (Gordon-Schowalter) derivative for a generic tensor $\boldsymbol{A}$ is defined as
\begin{equation}
\overset{\nabla}{\boldsymbol{A}}=\frac{\partial \boldsymbol{A}}{\partial t}+\boldsymbol{u}\cdot \nabla\boldsymbol{A}-\left(\nabla\boldsymbol{u}^T \cdot \boldsymbol{A}+\boldsymbol{A}\cdot\nabla\boldsymbol{u} \right).
\end{equation}

In the two-fluid miscible case, the scalar field $0 \le c \le 1$ defines the local concentration of the non-Newtonian fluid, found with the following advection-diffusion transport equation: 
\begin{equation}
\frac{\partial c}{\partial t}+\boldsymbol{u}\cdot \nabla c =D\nabla^2 c,
\label{eq:conc}
\end{equation}
where the parameter $D$ sets the diffusivity.
A complete formulation should also include the stress-induced diffusion of the polymer chains, \reva{$D\lambda/\eta_p \nabla\nabla$:$(\boldsymbol{\tau} c)$} \citep{apostolakis2002stress, dimitropoulos2006direct, vaithianathan2006improved}, but this term has often been neglected in advection-dominated problems \citep{dimitropoulos2006direct}, also owing to the numerical difficulties in its calculation \citep{vaithianathan2006improved, vaithianathan2007polymer, dimitropoulos2006direct}. The stress-induced diffusion usually becomes important at high values of the Deborah number \citep{apostolakis2002stress}, ratio of the elastic non-Newtonian time scale over a flow time scale. \reva{In this work we are in the limit of vanishing Deborah number; via dimensional analysis, it can be demonstrated that the contribution of the stress-induced diffusion is about one order of magnitude smaller than that from Fickian diffusion for the set of parameters selected in this work. As the contribution of stress-induced diffusion is of secondary importance (and it adds a significant computational overhead), it has been neglected in this work. }

The volume of fluid $0 \le \phi \le 1$ is adopted to track the local concentration of the non-Newtonian fraction for the two-fluid immiscible case, whose transport equation is:
\begin{equation}
\frac{\partial \phi}{\partial t} + \nabla\cdot(\boldsymbol{u}H) = 0.
\label{eq:vof}
\end{equation}
The cell-local color function $H$ defines the local concentration of the volume of fluid within a control volume $V$, equal to the computational cell. The local volume of fluid $\phi$ is obtained from the cell-local color function $H$ using the MTHINC (multi-dimensional tangent of hyperbola for interface capturing) method \citep{ii2012interface,RostiDB_2019}.
\revb{In the Navier-Stokes equations, Eq.~\ref{eq:ns}, no surface tension force is present: in all two-fluid immiscible cases the surface tension among the EVP and Newtonian phases is set to zero, hence surface tension forces are neglected. Within this setup the volume of fluid field can be understood as a concentration field at infinite Schmidt number (i.e. zero diffusivity, $Sc=\eta_s/(\rho D)$). }

The system of equations is discretized on a staggered, uniform, Cartesian grid, where pressure, density, viscosity, non-Newtonian extra-stresses, concentration and volume of fluid variables are stored at the cell center, while velocity data is stored at the cell faces. 
A second-order finite difference scheme is used to discretize spatial derivatives for all terms but the advection term in the extra-stress transport equation, for which a fifth-order WENO scheme is used \citep{shu2009high,sugiyama2011full}. 
The equations are advanced in time with a second-order, explicit Adams-Bashforth scheme. 
To enforce incompressibility, a fractional-step method \citep{kim1985application} is adopted, and a fast, Fourier-based pressure solver is used for the resulting Poisson equation for the pressure. 
The numerical solver is implemented on the in-house code \textit{Fujin}; see \url{https://groups.oist.jp/cffu/code} for additional details and validation tests. The very same code has been successfully used on problems involving Newtonian and non-Newtonian jets, EVP fluids and two-fluid problems \citep{abdelgawad2023scaling,cannon2021effect,cannon2024,soligo2023non}.

\section{Computational setup}
\label{sec: setup}

We perform numerical simulations of planar jets in a two-dimensional domain, with dimensions $L=L_x=L_y=200h$, with $h$ being the half-height of the inlet. The non-Newtonian EVP fluid is injected into the domain trough the inlet (of size $2h$) on the left side of the domain ($x=0$, $y=0$) into a pool of stagnant $a)$ EVP fluid (one-fluid case) or $b)$ Newtonian fluid (two-fluid miscible and immiscible cases).
An hyperbolic tangent profile \citep{stanley2002study,da2002influence} is imposed at the inlet,
\begin{equation}
u(0,y)=\frac{u_0+u_{\infty}}{2}-\frac{u_0-u_{\infty}}{2} \tanh{\left[\frac{h}{2\delta_\theta} \left(\frac{|y-L/2|}{h}-1 \right) \right]},
\label{eq:vin}
\end{equation}
with $u_0$ being the reference jet velocity, $u_\infty=0.1u_0$ a small co-flow velocity, and $\delta_\theta=h/30$ the momentum thickness of the shear-layer at the inlet.
In the results, we account for the effect of the co-flow by removing from the stream-wise velocity component $u$ the velocity at the top and bottom boundaries, \textit{i.e.} the stream-wise velocity presented in the following sections is $u(x,y)-(u(x,L/2)+u(x,-L/2))/2$ \citep{guimaraes2020direct}. 
At the inlet the polymers are not stretched ($\boldsymbol{\tau}=\boldsymbol{0}$). \revb{The effect of non-zero extra-stresses at the inlet has also been tested; we briefly report the main findings in appendix~\ref{app:inlet}.}
Free-slip conditions are imposed at the top and bottom boundaries, and a non-reflective outflow boundary condition \citep{orlanski1976simple} is imposed at the right boundary.

The Reynolds number at the inlet is set sufficiently low to be laminar for a Newtonian fluid \citep{deo2008influence, sureshkumar1995effect, sato1964experimental, soligo2023non}; in particular, we choose $Re=\rho h u_0 /\eta_t=20$, where $\eta_t=\eta_s+\eta_p$ is the total viscosity. For the Newtonian case $\eta_t=\eta_s$, while for the non-Newtonian cases we set the polymer viscosity equal to 10\% of the total viscosity, $\eta_p/\eta_t=0.1$. \revb{This value of the polymer viscosity corresponds to a dilute polymer concentration, and has been selected following previous numerical works \citep{abdelgawad2023scaling,le2020coherent,perlekar_manifestations_2006,perlekar2010direct,rosti2018turbulent}. }
Elastic turbulence can develop at this value of Reynolds numbers \citep{soligo2023non}, provided the Deborah number $De=\lambda u_0/h$ is large enough ($De \ge \mathcal{O}(10))$. In this work we choose a very small value of Deborah number $De=10^{-3}$, so that the planar jet is laminar, and thus the simulations can be limited to a two-dimensional domain, the stress-induced polymer diffusion is negligible in Eq.~\ref{eq:conc}, and all the different dynamics can be attributed to plasticity alone.

The Bingham number is defined as the ratio of the yield stress $\tau_y$ of the EVP fluid over a characteristic stress; we define the Bingham number as $Bi=\tau_y h /(\eta_t u_0)$. In this work, we are interested in the effect of the Bingham number on the structure and dynamics of the jets. Thus, we vary the Bingham number from $Bi=0$ (the material is always yielded) up to $Bi=1$, for which we observe large portions of unyielded fluid. \reva{With the present configuration, the ratio of Bingham over Reynolds is always $Bi/Re\ll1$, with the largest value attained at $Bi=1$, $Bi/Re=0.05$. }
Additional parameters need to be set for the two-fluid cases: the diffusivity of the concentration field (miscible case) and the surface tension of the interface (immiscible case). The Schmidt number, defining the ratio of momentum over mass diffusivity, has been set to $Sc=\eta_s/(\rho D)=1$ in all cases, where we defined the Schmidt number with the solvent viscosity, $\eta_s=0.9\eta_t$. For the immiscible case we consider an extremely deformable interface, thus setting the Weber number $We=\rho u_0^2 h/\sigma =\infty$, with $\sigma$ being the surface tension of the interface. \revb{With this choice of surface tension, the volume of fluid can be understood as a concentration field having an infinite Schmidt number. Furthermore, in the present configuration (laminar planar jet) we expect the contribution of surface tension forces to be negligible nonetheless: the curvature of the volume of fluid interface is nearly zero, exception made for the region close to the inlet.}

The list of all simulated cases is reported in Tab.~\ref{tab:cases}; five different values of the Bingham number are considered for each non-Newtonian case. The different configurations are: $(i)$ EVP jet issued into EVP ambient fluid (EVP-EVP, EE), $(ii)$ EVP jet issued into Newtonian ambient fluid, miscible case (EVP-Newtonian - miscible, EN-m), and $(iii)$ EVP jet issued into Newtonian ambient fluid, immiscible case (EVP-Newtonian - immiscible, EN-i). A reference Newtonian case with a Newtonian jet issued into a Newtonian ambient fluid is also simulated (Newtonian-Newtonian, NN). In this latter case the concentration and volume of fluid fields are included within the same simulation, as there is no back reaction onto the flow.

\begin{table}[width=.9\linewidth,cols=3,pos=h]
\caption{List of simulations; the case column defines the configuration, as in jet-ambient fluid (and miscible/immiscible).}
\label{tab:cases}
\begin{tabular*}{\tblwidth}{@{} LCC@{} }
\toprule
Case & Legend entry  & $Bi$ \\
\midrule
Newtonian-Newtonian & NN ~ \color[HTML]{888888}\tiny{$\pentagofill$} & - \\
\midrule
EVP-EVP & EE ~ \color[HTML]{ff0000}$\circ$ & $0$ \\
EVP-EVP & EE ~ \color[HTML]{ff4000}$\circ$ & $10^{-3}$ \\
EVP-EVP & EE ~ \color[HTML]{ff8000}$\circ$ & $10^{-2}$ \\
EVP-EVP & EE ~ \color[HTML]{ffbf00}$\circ$ & $10^{-1}$ \\
EVP-EVP & EE ~ \color[HTML]{ffff00}$\circ$ & $1$ \\
\midrule
EVP-Newtonian - miscible & EN-m ~ \color[HTML]{0000ff}$\diamondsuit$ & $0$ \\
EVP-Newtonian - miscible & EN-m ~ \color[HTML]{0040df}$\diamondsuit$ & $10^{-3}$ \\
EVP-Newtonian - miscible & EN-m ~ \color[HTML]{0080bf}$\diamondsuit$ & $10^{-2}$ \\
EVP-Newtonian - miscible & EN-m ~ \color[HTML]{00bf9f}$\diamondsuit$ & $10^{-1}$ \\
EVP-Newtonian - miscible & EN-m ~ \color[HTML]{00ff80}$\diamondsuit$ & $1$ \\
\midrule
EVP-Newtonian - immiscible & EN-i ~ \color[HTML]{000000}\tiny{$\square$} & $0$ \\
EVP-Newtonian - immiscible & EN-i ~ \color[HTML]{4f3220}\tiny{$\square$} & $10^{-3}$ \\
EVP-Newtonian - immiscible & EN-i ~ \color[HTML]{9d643f}\tiny{$\square$} & $10^{-2}$ \\
EVP-Newtonian - immiscible & EN-i ~ \color[HTML]{ec955f}\tiny{$\square$} & $10^{-1}$ \\
EVP-Newtonian - immiscible & EN-i ~ \color[HTML]{ffc77f}\tiny{$\square$} & $1$ \\
\bottomrule
\end{tabular*}
\end{table}

\begin{figure}[ht!]
\centering
\includegraphics[width=0.9\columnwidth]{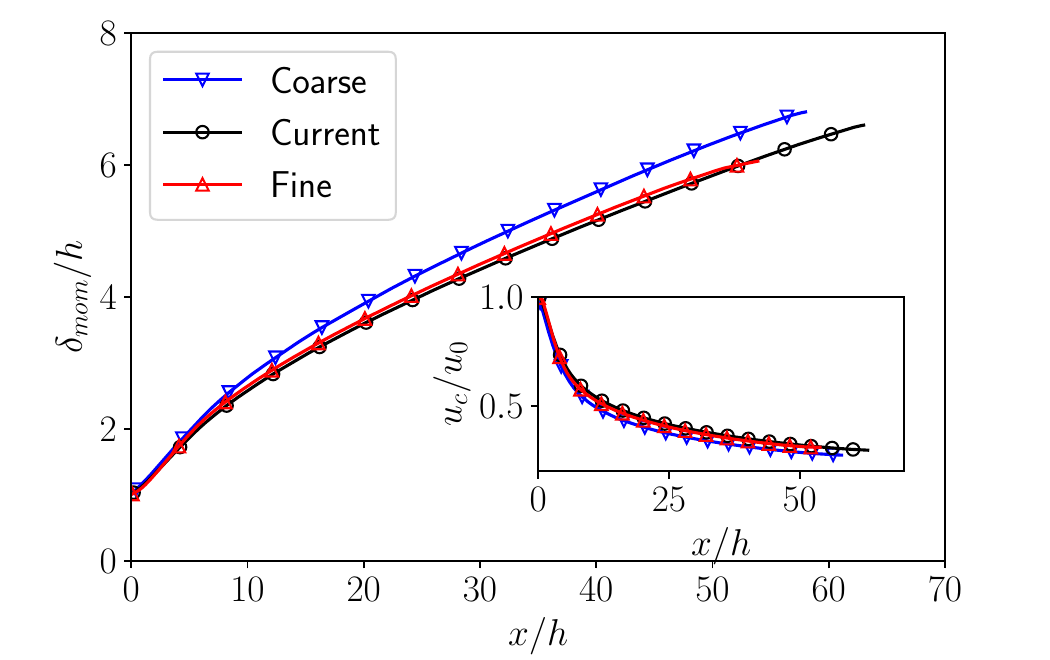}
\caption{Jet thickness and centerline velocity for the EN-i case at Bingham $Bi=1$ for three different grid resolutions: coarse ($500$ grid points), current ($1000$ grid points) and fine ($2000$ grid points). Data are plotted along the stream-wise direction as long as the fluid at the centerline is yielded.}
\label{fig: gridres}
\end{figure}

The computational domain is discretized using $N=N_x=N_y=1000$ grid points in both the stream-wise and the jet-normal directions. 
To verify the grid independence of our results, we performed two additional simulations for the EN-i $Bi=1$ case, one at a lower grid resolution ($500$ grid points, $0.5\times$ the baseline resolution) and one at a higher grid resolution ($2000$ grid points, $2\times$ the baseline resolution).
Note that the immiscible EVP-Newtonian case at the highest Bingham number (EN-i, $Bi=1$) was selected for the grid independence test as it is the most computationally challenging case: indeed, $(i)$ among all configurations it has the thinnest jet core, $(ii)$ strong non-Newtonian extra-stress gradients are present at the interface between the EVP jet fluid and the Newtonian ambient fluid, and $(iii)$ a large unyielded plug spans a large fraction of the jet core. The jet thickness $\delta_{mom}$ and the centerline velocity $u_c$ (see section~\ref{sec: results} for the definition of these quantities) are reported in figure~\ref{fig: gridres} for the three grid resolutions tested in this work. Data clearly shows that the grid resolution adopted in the present work, $1000$ grid points, is sufficient to capture all flow dynamics of the jet; doubling the grid resolution, $2000$ grid points, does not significantly change the jet statistics. On the other hand, a lower grid resolution, $500$ grid points, leads to a higher spreading rate of the jet thickness and a higher decay rate of the centerline velocity. 



\section{Results}
\label{sec: results}

\begin{figure*}[ht!]
\centering
\begin{tikzpicture}
    \node at (0,0) {\includegraphics[width=0.9\textwidth]{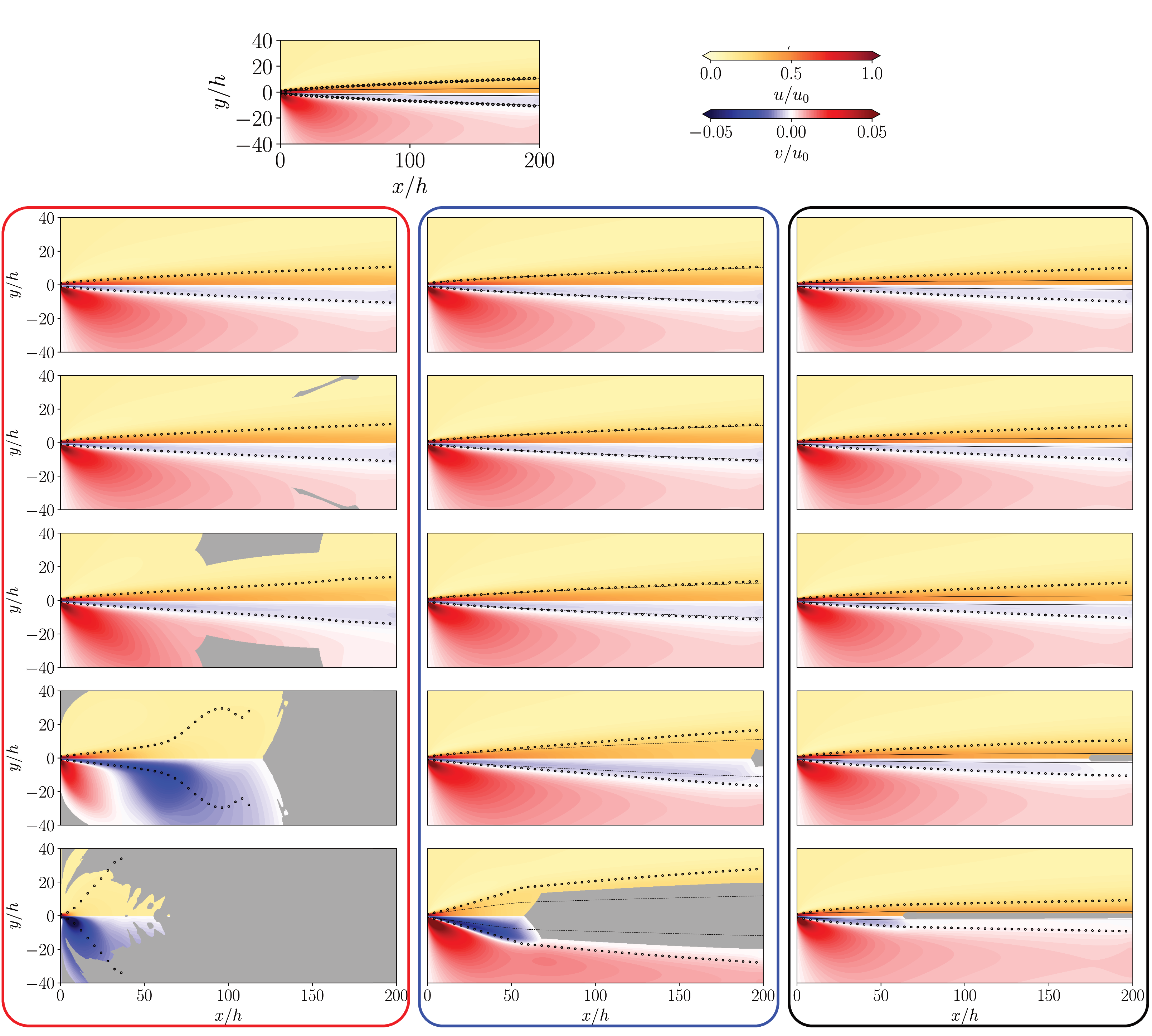}};
    \node at (-5,6.35) {$(a)$};
    \node at (-6.7,3.7) {$(b)$};
    \node at (-1.7,3.7) {$(g)$};
    \node at (3.3,3.7) {$(l)$};
    \node at (-6.7,1.6) {$(c)$};
    \node at (-1.7,1.6) {$(h)$};
    \node at (3.3,1.6) {$(m)$};
    \node at (-6.7,-0.5) {$(d)$};
    \node at (-1.7,-0.5) {$(i)$};
    \node at (3.3,-0.5) {$(n)$};
    \node at (-6.7,-2.6) {$(e)$};
    \node at (-1.7,-2.6) {$(j)$};
    \node at (3.3,-2.6) {$(o)$};
    \node at (-6.7,-4.7) {$(f)$};
    \node at (-1.7,-4.7) {$(k)$};
    \node at (3.3,-4.7) {$(p)$};
\end{tikzpicture}
\caption{Flow visualization of the planar jet; the top and bottom parts of the domain are omitted for clarity (far-field in the jet-normal direction). The NN case is shown on top, whereas colored boxes group the different cases: EE (red box), EN-m (blue box) and EN-i (black box). In each box, the Bingham number increases from top ($Bi=0$) to bottom ($Bi=1$). In each panel, the top half shows the stream-wise velocity $u$ and the bottom half shows the jet-normal velocity $v$; gray shading identifies unyielded regions. A solid black line identifies the thickness of the volume of fluid, $\delta_\phi$, a dashed black line identifies the thickness of the concentration field, $\delta_c$, and circle black markers identify the jet thickness, $\delta_{mom}$. }
\label{fig: qualiuv}
\end{figure*}
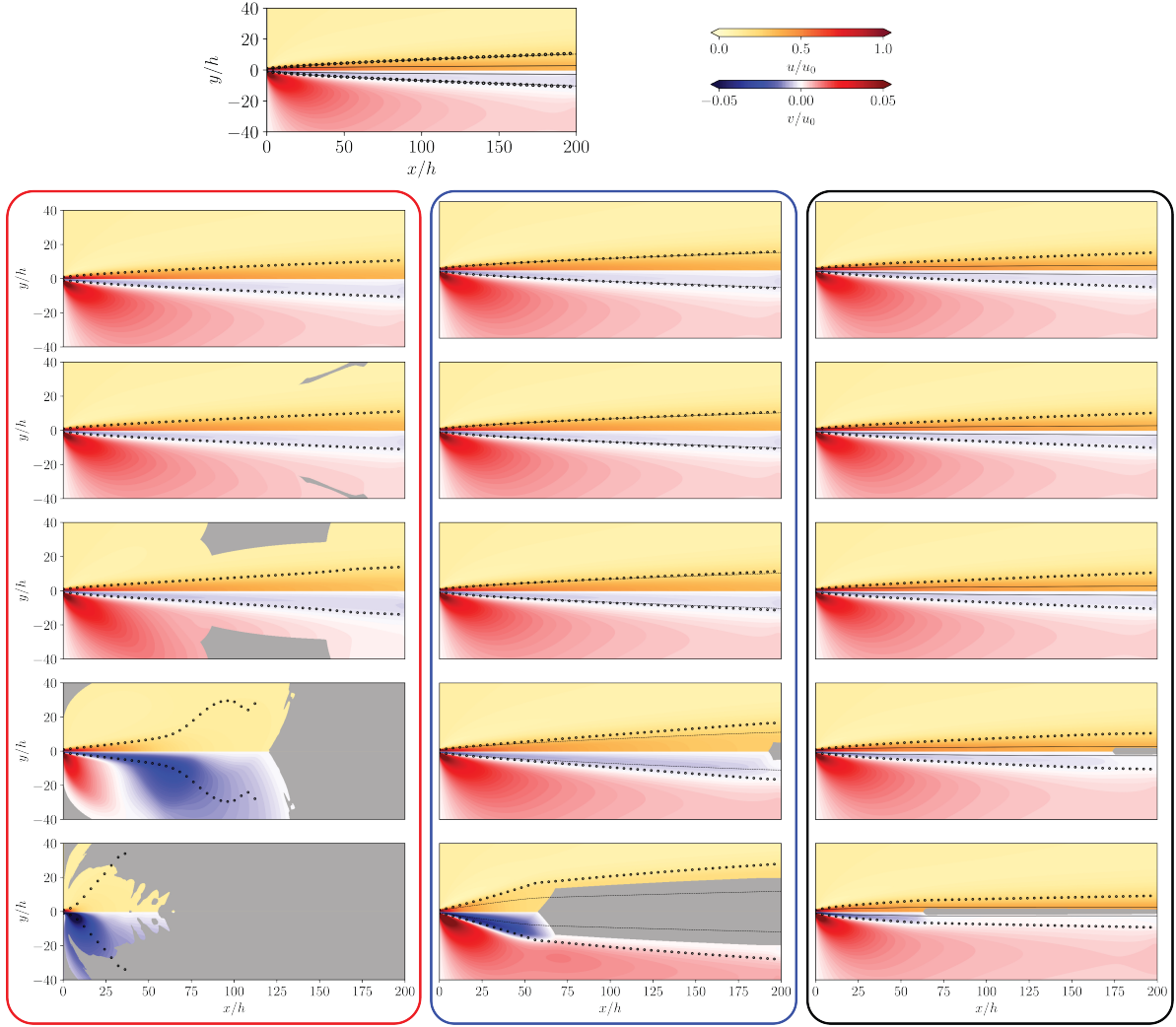

We report in figure~\ref{fig: qualiuv} a visualization of the flow field, with the stream-wise velocity $u$ shown in the top half of each panel and the jet-normal velocity $v$ shown in the bottom half. The Newtonian case, NN, is reported \reva{panel $(a)$}, whereas a colored box identifies each configuration: EVP-EVP (red, \reva{panels $(b)$-$(f)$}), EVP-Newtonian - miscible (blue, \reva{panels $(g)$-$(k)$}), and EVP-Newtonian - immiscible (black, \reva{panels $(l)$-$(p)$}). For the non-Newtonian cases, each row corresponds to a different Bingham number, increasing from top ($Bi=0$) to bottom ($Bi=1$). Black circle markers show the jet thickness, $\delta_{mom}$, defined as the half-width of an inviscid, shearless flow with velocity equal to the centerline velocity $u_c(x)$ and carrying the same volumetric flow-rate of the jet,
\begin{equation}
\delta_{mom}(x)=\frac{1}{2}\frac{\int_{-\infty}^{+\infty} u(x,y) \text{d}y}{u_c(x)}.
\label{eq:delta}
\end{equation}
The thickness of the scalar field is also reported with solid ($\delta_\phi$, volume of fluid) and dashed ($\delta_c$, local concentration) lines. The thickness is defined as the distance from the centerline at which the local value of the scalar field (the volume of fluid or the local concentration) equals half of its value at the centerline. For the immiscible EVP-Newtonian cases this corresponds to the position of the interface between the two fluids. Note that, we report both $\delta_\phi$ and $\delta_c$ for the NN case, since in this case the volume of fluid and local concentration fields are passive, and they act as simple markers to distinguish the jet from the ambient fluid.
A grey shaded area identifies the unyielded fluid; to distinguish the yielded and unyielded regions we use the yield variable $Y=\max\left(0, 1-\alpha\tau_y/|\tau_d|\right)\in [0,1]$ from the Saramito elastoviscoplastic fluid model. 
In this work, we set the threshold value to $Y=10^{-4}$ for all cases.

For the EVP-EVP (EE) cases, at the lowest Bingham number, $Bi=0$, there is no difference with the NN case: the Deborah number used in this study is very small, hence elastic effects are negligible, and the fluid is yielded in the entire domain.
As the Bingham number is increased we observe the formation of unyielded regions, as indicated by the grey shaded area; at low Bingham numbers, $Bi=10^{-3}$ and $Bi=10^{-2}$, the regions of unyielded fluid are limited to the ambient fluid. 
At larger values of the Bingham number, $Bi=10^{-1}$ and $Bi=1$, a substantial portion of the fluid becomes unyielded: the region close to the inlet is yielded whereas away from the jet centerline and near the outlet the fluid is unyielded. The formation of the unyielded  plug increases the spreading rate of the jet. \reva{At high Bingham number, the shape of the yielded fluid may be reminiscent of a Saffman-Taylor instability, with finger-like structures extending into the unyielded fluid, as observed in previous studies \citep{lindner2000viscous,maleki2005saffman,ebrahimi2016viscous,eslami2017viscous,eslami2020viscoplastic}. This however is not the case here for the EE configuration, as a key ingredient for the Saffman-Taylor instability is missing: the EE configuration is a characterized by a single EVP fluid phase and no interface is present. The finger-like pattern we observed is merely an effect of the yield variable.}

For the miscible EVP-Newtonian (EN-m) cases, we do not observe any appreciable change from the NN reference case as long as the jet fluid is yielded, up to $Bi=10^{-2}$ in this work. As the ambient fluid is Newtonian, there are no unyielded regions in the ambient fluid, as observed instead for the EE cases.
The formation of an unyielded plug in the core of the jet is observed beyond Bingham numbers $Bi\ge10^{-1}$; the unyielded plug is limited to the core of the jet near the outlet at $Bi=10^{-1}$ and it extends beyond $\delta_c$ and further towards the inlet at $Bi=1$. The EVP and Newtonian fluids are miscible in this configuration, hence it is expected that the unyielded region extends beyond the concentration thickness, $\delta_c$. An increase of the spreading rate of the jet is observed upstream of the unyielded plug: the presence of the plug pushes the injected fluid away from the centerline of the jet, as indicated by the larger magnitude of the jet-normal velocity at $Bi=1$.

A similar result is reported for the immiscible EVP-Newtonian (EN-i) cases. At values of the Bingham number smaller than $Bi=10^{-3}$, the fluid is yielded everywhere and no differences from the NN case are observed, as expected from the negligible contribution of fluid elasticity. 
An unyielded plug is reported for values of the Bingham number larger than $Bi\ge10^{-2}$: the unyielded fluid plug forms within the volume of fluid phase near the outlet and further extends towards the inlet as the Bingham number is increased. A thin region of yielded EVP fluid is present in between the unyielded plug and the Newtonian ambient fluid. The presence of a plug of unyielded fluid has little effect on the jet flow, with only a minor increase in the magnitude of the jet-normal velocity observed upstream of the unyielded plug. Compared to the miscible EVP-Newtonian case, we observe a much smaller size of the unyielded plug and a lesser blockage effect.

\begin{figure}[ht!]
\centering
\begin{tikzpicture}
    \node at (0,14.1) {\includegraphics[width=0.9\columnwidth]{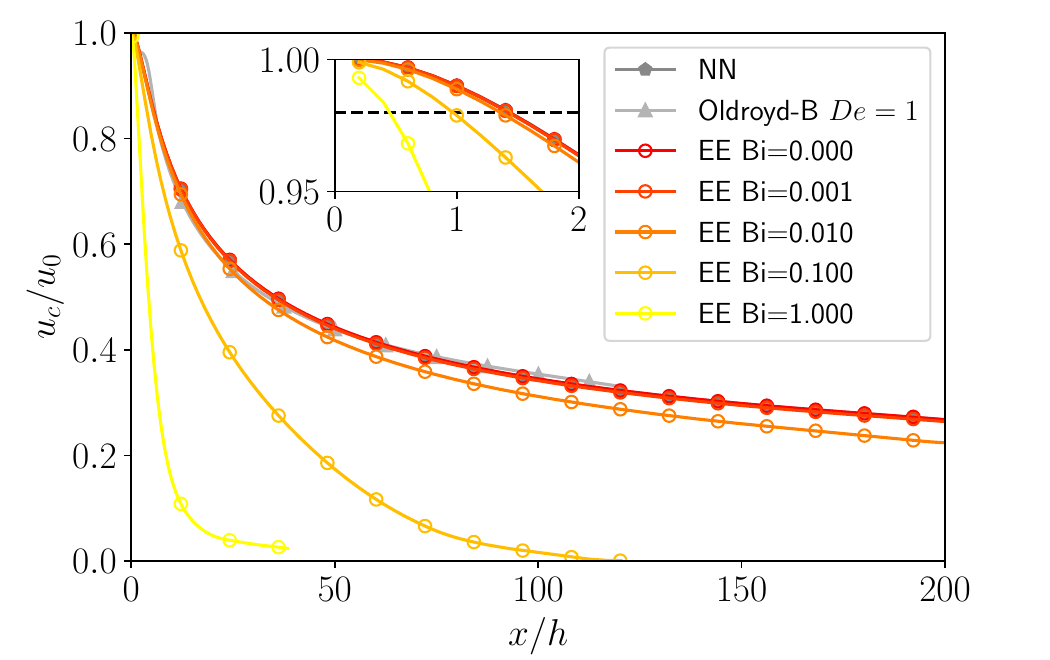}};
    \node at (0,9.4) {\includegraphics[width=0.9\columnwidth]{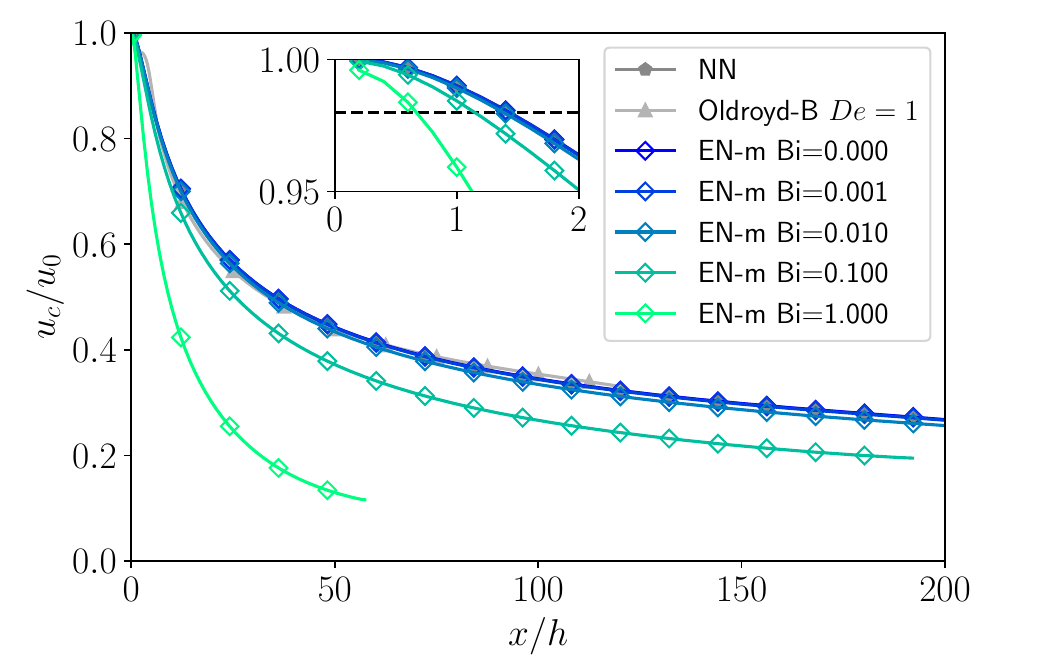}};
    \node at (0,4.7) {\includegraphics[width=0.9\columnwidth]{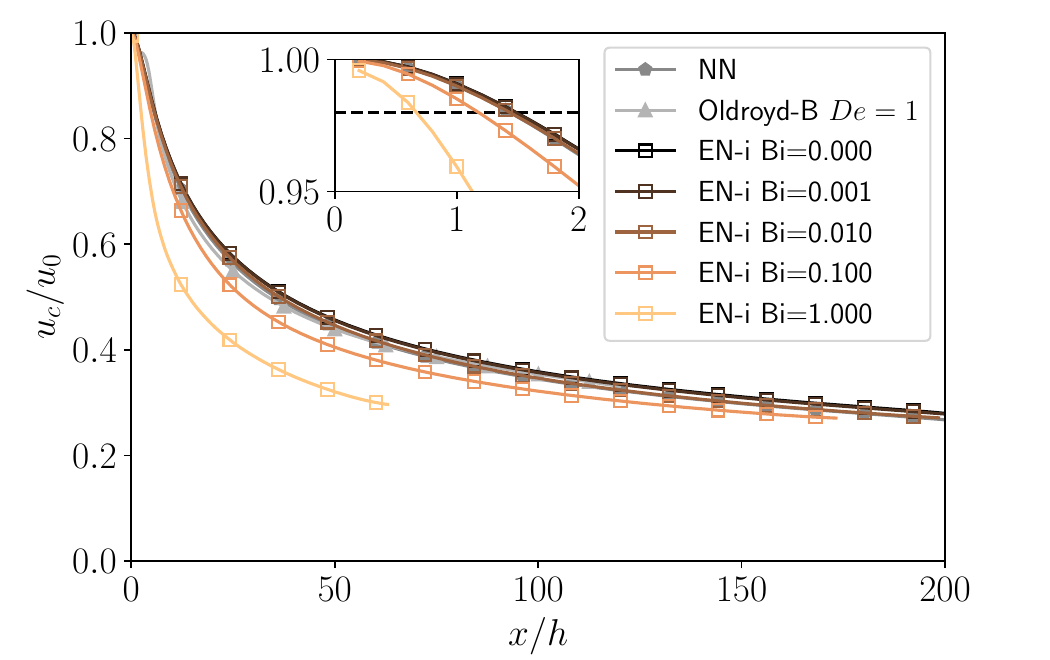}};
    \node at (0,0.0) {\includegraphics[width=0.9\columnwidth]{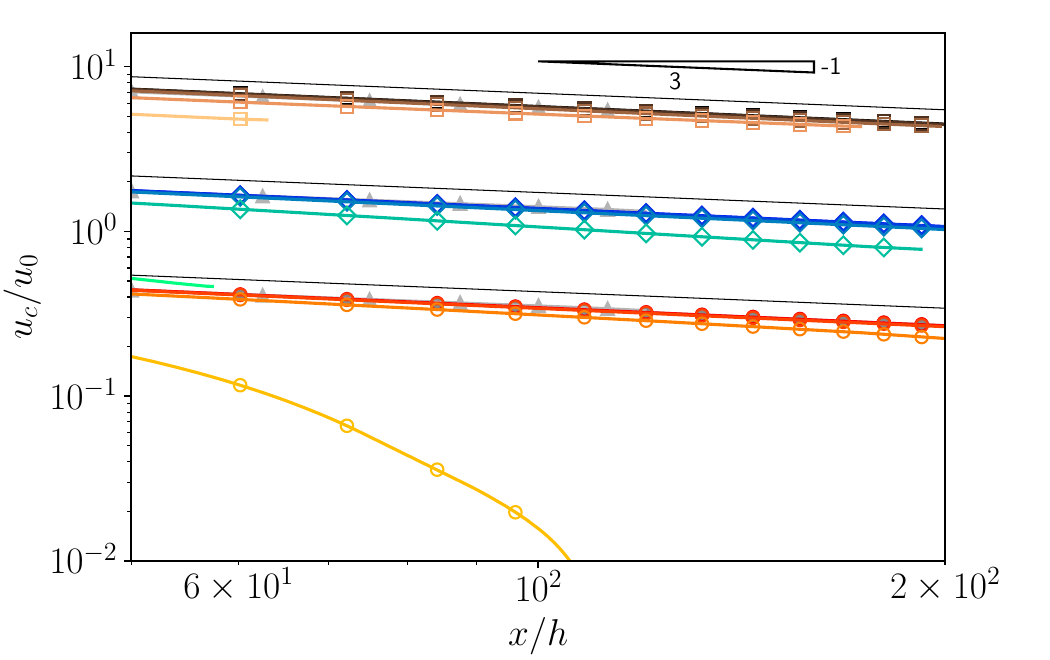}};
    \node at (-3.6,16.) {$(a)$};
    \node at (-3.6,11.3) {$(b)$};
    \node at (-3.6,6.4) {$(c)$};
    \node at (-3.6,1.7) {$(d)$};
\end{tikzpicture}
\caption{Centerline velocity of the jet. Each panel refers to a different configuration \reva{(panel $(a)$: EE; panel $(b)$: EN-m; panel $(c)$: EN-i), whereas panel $(d)$} reports all data in logarithmic scale to highlight the laminar scaling, \reva{together with the $u_c/u_0\propto (x/h)^{-1/3}$ analytic scaling (black thin lines)}. In \reva{panel $(d)$}, data are shifted \reva{upwards} for each configuration for clarity. The NN case is reported as a reference for all configurations. Data are plotted along the stream-wise direction as long as the fluid at the centerline is yielded. \reva{Data from \citep{soligo2023non} for the Oldroyd-B fluid at $De=1$ is reported for reference.}}
\label{fig: vc}
\end{figure}

The stream-wise velocity along the centerline of the jet, $u_c$, is reported in figure~\ref{fig: vc}. Each panel refers to a different configuration (EE \reva{in panel $(a)$}, EN-m \reva{in panel $(b)$}, EN-i \reva{in panel $(c)$}), whereas panel \reva{$(d)$} reports in logarithmic scale all data (shifted \reva{upwards} by configuration for ease of reading) to highlight the far-field power-law scaling. The centerline velocity is reported only until the fluid becomes unyielded at the centerline. \reva{The centerline velocity for an Oldroyd-B fluid at $De=1$ (see the simulation database in \citet{soligo2023non}) is reported for reference. This latter case is representative of a fully-yielded EVP fluid (\citet{saramito2007new} model), although at a much larger Deborah number. }
For all configurations we observe a faster decay of the centerline velocity for increasing values of the Bingham number: as the Bingham number is increased a larger fraction of the fluid becomes unyielded and impedes the flow of the jet.
There is a clear trend based on the configuration: the EE configuration has the highest decay rate, followed by the EN-m and lastly by the EN-i, which has the lowest decay rate.
The characteristics of the ambient fluid explain the trend: for the EE configuration the ambient fluid is EVP fluid, same as the jet fluid, thus it becomes unyielded even at relatively low values of the Bingham number due to the low magnitude of the deviatoric stress. The unyielded fluid acts as a blockage and causes a higher decay rate of the centerline velocity for increasing values of the Bingham number, as larger portions of fluid become unyielded.
In the EN-m and EN-i configurations the EVP fluid, and thus the unyielded region, is limited to the jet and near-field (miscible case). 
The mixing of EVP jet fluid and Newtonian ambient fluid in the EN-m configuration limits the presence of EVP fluid to the jet and near-field of the jet: as shown in figure~\ref{fig: qualiuv}, the unyielded fluid at high Bingham partially extends beyond $\delta_c$ (identified by a dashed line). The presence of an extensive unyielded plug causes a noticeable widening of the jet upstream of the blockage. 
In the immiscible configuration the presence of EVP fluid is limited to the jet, \textit{i.e.} within $\delta_\phi$. 
At large values of the Bingham number a viscoelastic plug is formed within the jet fluid close to the outlet and extends further towards the inlet as the Bingham number is increased; a thin layer of yielded EVP fluid is observed in between the unyielded plug and the ambient Newtonian fluid. 
The presence of the plug has relatively little effect on the decay rate of the centerline velocity, which becomes clear only at the highest Bingham number tested. For the immiscible configuration, the low thickness of the unyielded plug causes a minor blockage of the jet flow: hence, we do observe a limited change in the jet thickness and in the decay rate of the centerline velocity as the unyielded plug is pushed out. 
It is interesting to note that for the EE and EN-m configurations up to $Bi=10^{-3}$ the centerline velocity overlaps with the NN configuration reasonably well, and the centerline velocity decay rate increases only for larger Bingham numbers. 
For the immiscible configuration (EN-i) instead, we observe that the centerline velocity of the case at $Bi=10^{-2}$ overlaps with the data for the NN configuration; a slightly slower decay compared to the NN configuration is reported for smaller values of the Bingham number. 

\reva{Figure~\ref{fig: vc}$(d)$} shows the centerline velocity data for all cases in logarithmic scale; data for different configurations is shifted vertically for clarity of reading. 
It has been demonstrated that the centerline velocity of a laminar, Newtonian planar jet scales with the stream-wise coordinate as $u_c/u_0\propto (x/h)^{-1/3}$ \citep{Mei_2001}; this scaling has been found to hold true also for laminar, non-Newtonian planar jets \citep{parvar2020local,soligo2023non}.
The EVP fluid model we adopt behaves as an Oldroyd-B fluid when yielded and as a Kelvin-Voigt fluid when unyielded; given the very low contribution of elasticity ($De=10^{-3}$) adopted in the current setup, we expect good agreement with the Newtonian laminar scaling as long as the fluid is yielded. Indeed, all cases at low Bingham number show a power-law scaling $u_c/u_0\propto (x/h)^{-1/3}$ away from the inlet, while for large Bingham number the scaling starts to differ once the unyielded region alters the flow dynamics. 

\begin{figure}[ht!]
\centering
\includegraphics[width=0.9\columnwidth]{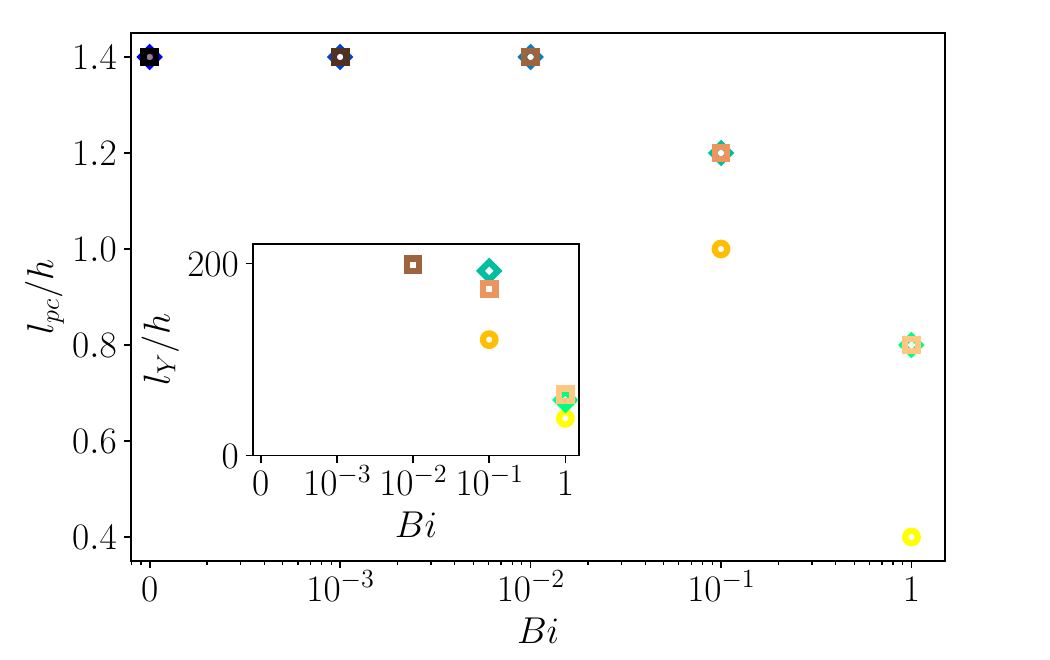}
\caption{\reva{Potential core length, $l_{pc}$, defined as the stream-wise distance where the centerline velocity equals $0.98 u_0$. In the inset we show the yield length, i.e., the distance from the inlet $l_Y$ at which unyielded fluid is first found at the centerline (unyielded fluid at the inlet is not considered in this measure).} Circle markers identify the EE configuration, diamonds markers the EN-m configuration and square markers the EN-i configuration; colors and markers for each case are the same as reported in table~\ref{tab:cases}.}
\label{fig:potcore}
\end{figure}

The potential core is a region near the inlet where the centerline velocity of the jet is nearly constant, and it is commonly defined as up to the stream-wise coordinate where the centerline velocity is equal to 98\% of the centerline velocity at the inlet \citep{deo2008influence}. The effect of the Bingham number on the length of the potential core is highlighted in the insets of figure~\ref{fig: vc} \reva{and in figure~\ref{fig:potcore}}. 
We report a general trend of decreasing length of the potential core for increasing values of the Bingham number for all configurations. A faster decay rate of the potential core is reported for the EE configuration, whereas we report a similar decay rate for both EVP-Newtonian configurations (miscible and immiscible). As discussed before, the higher decay rate for the EE configuration owes to the presence of extensive regions of unyielded fluid in both the jet and ambient fluid, leading to strong blockage. 
For the EVP jet in Newtonian ambient fluid, the unyielded fluid is limited to the jet region and the unyielded fluid region has a minor effect on the potential core length. 
\reva{In the inset of figure~\ref{fig:potcore} we report the yield length $l_Y$, which is defined as the stream-wise distance from the inlet at which unyielded fluid is first found. To compute $l_Y$ the unyielded fluid that is present at the inlet is not considered; if a data point is not reported the fluid at the centerline is yielded within the computational domain. As the Bingham number is increased, $l_Y$ reduces as unyielded fluid is found closer to the inlet. For laminar jets, the yield length can be linked to the laminar length: the laminar length in fully-laminar jets is defined as the distance from the inlet where the jet completely diffuses in the ambient flow \citep{kumar1984laminar}. In the experimental measurements of \citet{kumar1984laminar} a birefringent fluid was used and, for fully-laminar jets, the laminar length was defined as the distance at which the fringe pattern was no longer observed, i.e. where the shear stresses were so small that the working fluid was nearly optically isotropic. Here, to distinguish yielded and unyielded fluid we adopt a criterion based on the yield variable, namely the fluid is considered unyielded for $Y \le 10^{-4}$. The yield length is thus based as well on a similar concept as the laminar length: a stress-based threshold. At low values of the Reynolds number ($50\le Re\le200$) \citet{kumar1984laminar} report an increase in the laminar length of the jet with the Reynolds number. In this work we do not investigate the effect of the Reynolds number; we report instead a decrease in the yield length for increasing values of the Bingham number. }



\begin{figure}[ht!]
\centering
\begin{tikzpicture}
    \node at (0,9.4) {\includegraphics[width=0.9\columnwidth]{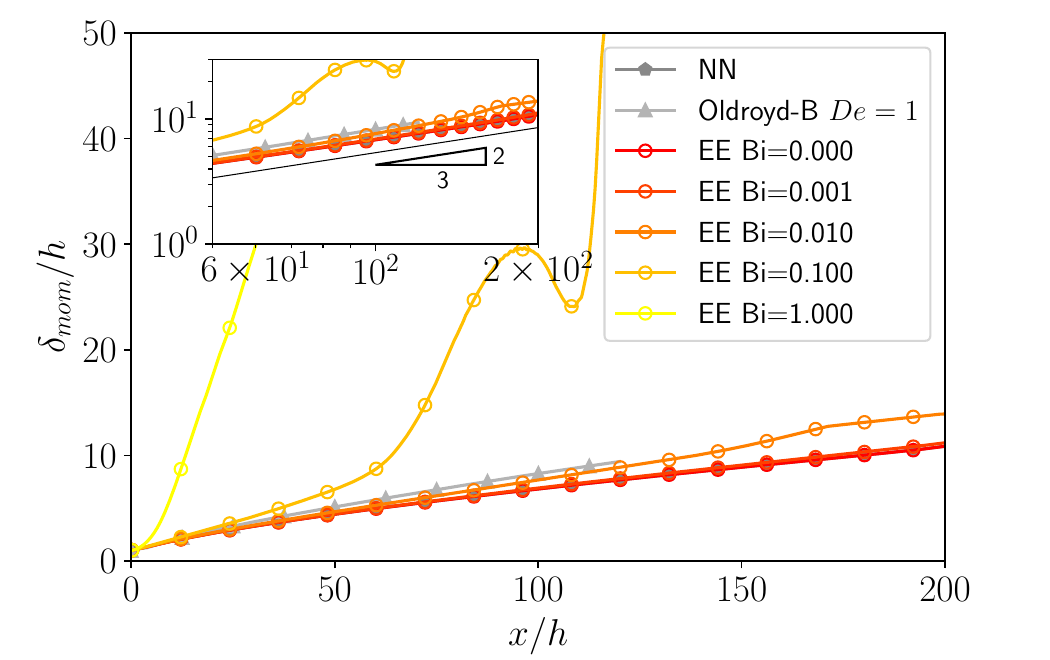}};
    \node at (0,4.7) {\includegraphics[width=0.9\columnwidth]{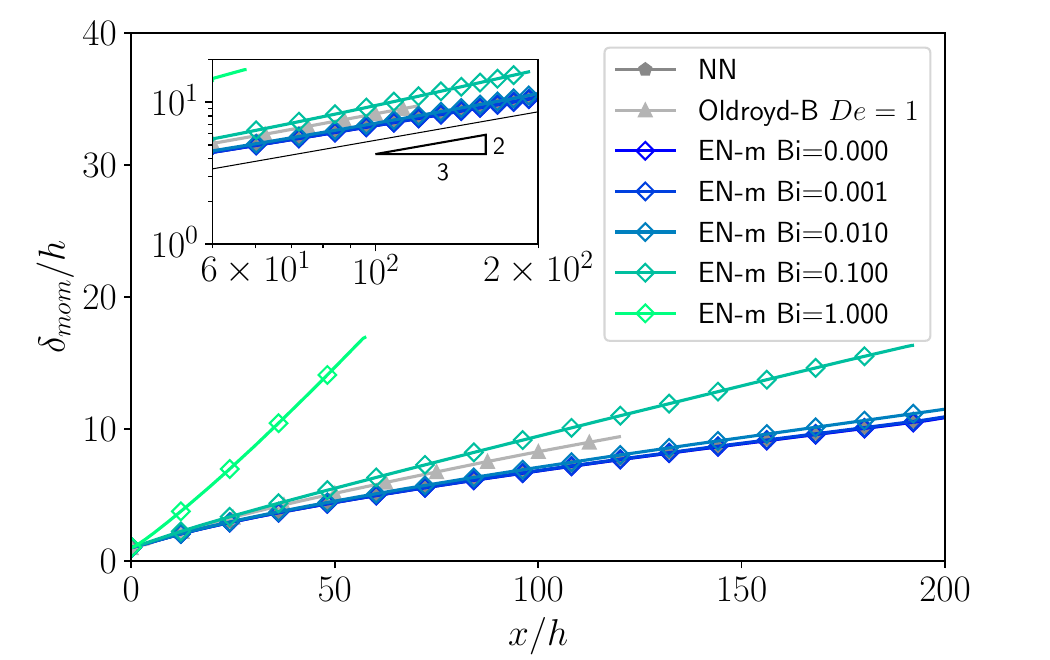}};
    \node at (0,0.0) {\includegraphics[width=0.9\columnwidth]{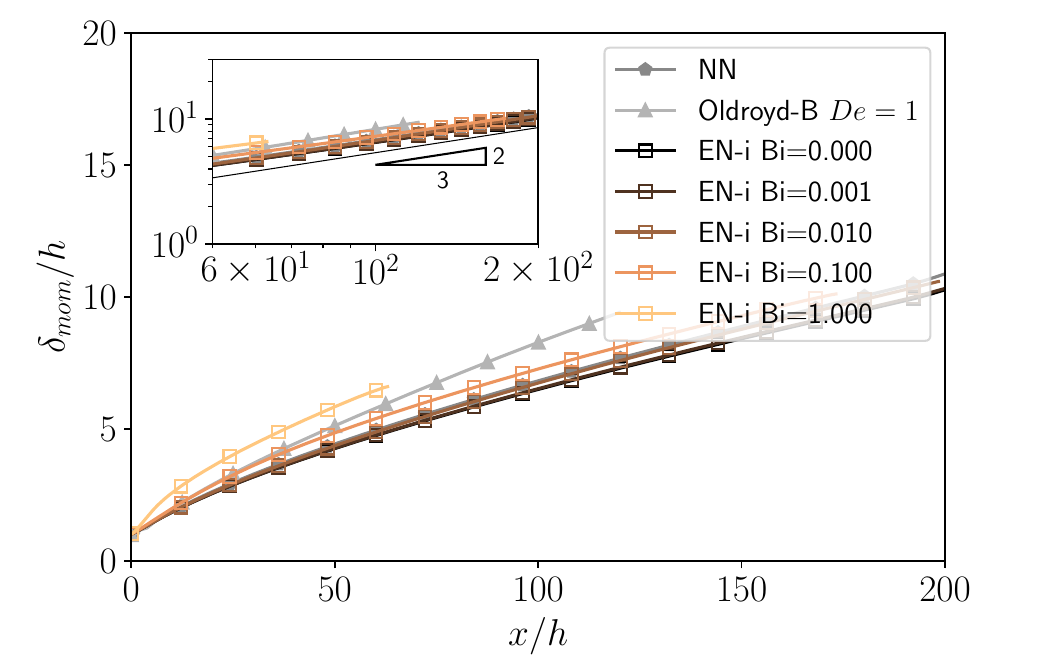}};
    \node at (-3.6,11.3) {$(a)$};
    \node at (-3.6,6.4) {$(b)$};
    \node at (-3.6,1.7) {$(c)$};
\end{tikzpicture}
\caption{Momentum thickness of the jet, $\delta_{mom}$, for the different configurations. The inset shows the jet thickness $\delta_{mom}$ in logarithmic scale for each configuration, together with the expected scaling for Newtonian, laminar, planar jets \reva{(black thin lines)}.}
\label{fig: delta}
\end{figure}

The thickness of the jet $\delta_{mom}$, is reported in figure~\ref{fig: delta}\reva{; we report as well data for the jet thickness of an Oldroyd-B jet in a similar configuration at $De=1$ from the simulation database of \citet{soligo2023non}. }
We also computed the jet thickness $\delta_{0.5}$, defined as the distance from the centerline at which the stream-wise velocity equals half the centerline velocity $u_c$ and noticed no appreciable changes in the trends between $\delta_{mom}$ and $\delta_{0.5}$, with only minor differences in the actual values. 
The inset within each panel repeats the $\delta_{mom}$ data in logarithmic scale to better show the presence of a power-law scaling\reva{; thin solid black lines show the power-law scaling}. 
We note that the jet thickness increases with the Bingham number for all the configurations. The effect of the Bingham number is different among the various configurations: similarly to what observed for the centerline velocity, the jet thickness for the EE configuration has the largest change with growing Bingham numbers, followed by the EN-m and then by the EN-i. 
At very low Bingham number the EE and EN-m configurations have a jet thickness that overlaps with that of the NN configuration: the majority of the fluid is unyielded and the effect of fluid elasticity is negligible 
The jet thickness for laminar, Newtonian planar jets has been shown to scale with the stream-wise coordinate as $\delta_{mom}/h\propto (x/h)^{2/3}$ \citep{Mei_2001}, and this scaling has been extended to laminar, non-Newtonian planar jets as well \citep{parvar2020local,soligo2023non}. 
The data in the inset of the plots for the EE and EN-m configurations clearly indicate that, at low Bingham numbers, the jet thickness follows the power-law scaling calculated for laminar, Newtonian, planar jets. Deviations from the scaling are observed in conjunction with the presence of large regions of unyielded fluid only. 
The EN-i configuration follows the power-law scaling $(x/h)^{2/3}$ up to $Bi=10^{-1}$; at very low Bingham numbers, smaller than $Bi<10^{-2}$, we observe a smaller jet thickness compared to the NN configuration. 


\begin{figure}[ht!]
\centering
\begin{tikzpicture}
    \node at (0,4.7) {\includegraphics[width=0.9\columnwidth]{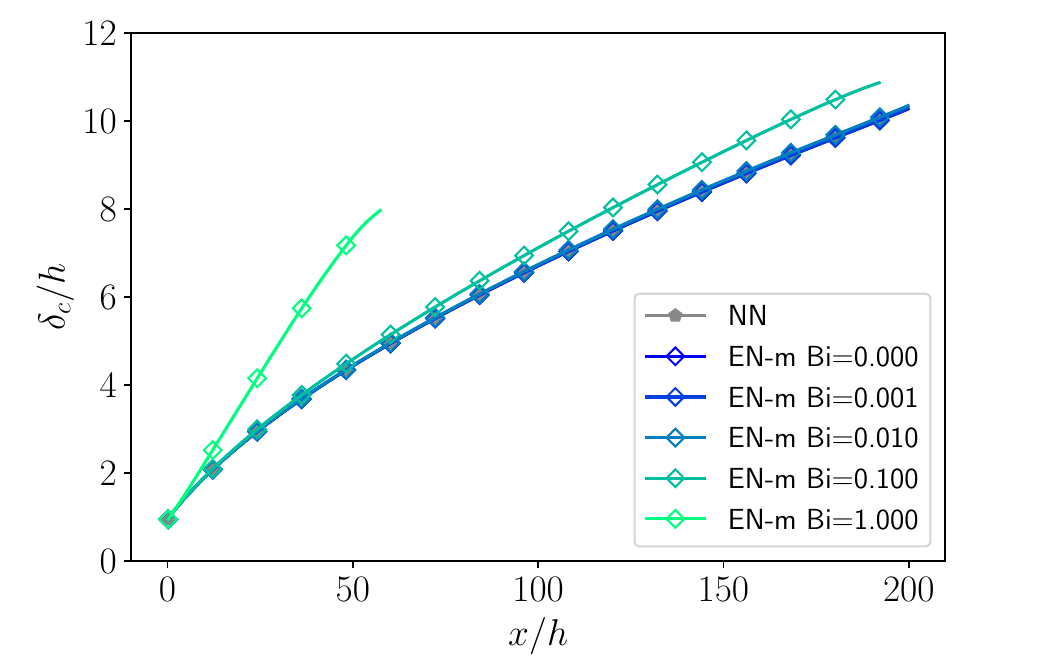}};
    \node at (0,0.0) {\includegraphics[width=0.9\columnwidth]{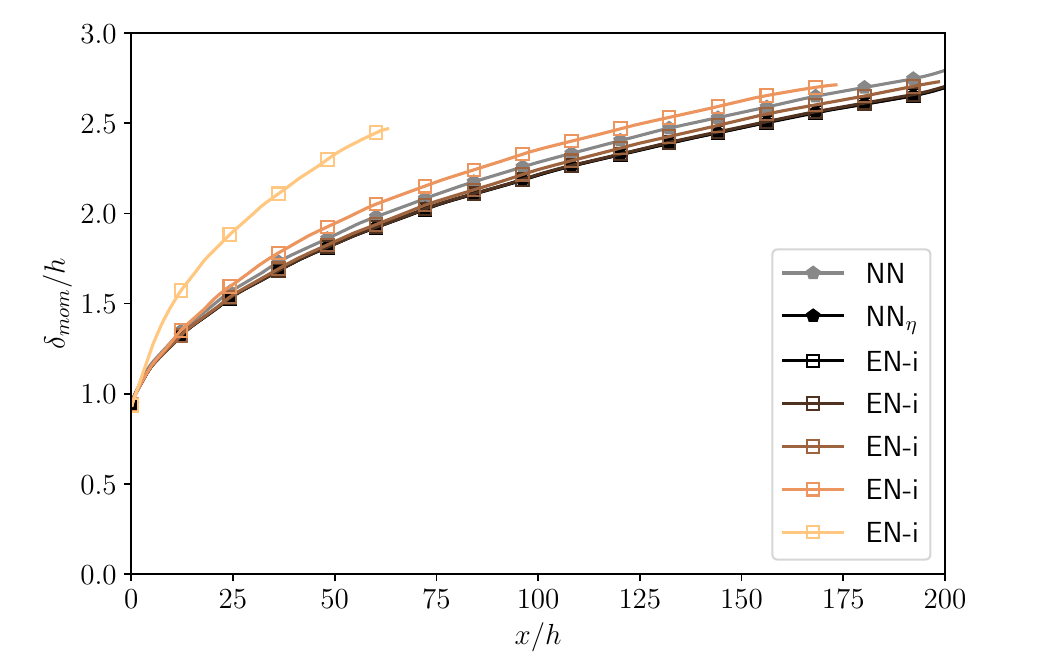}};
    \node at (-3.6,6.4) {$(a)$};
    \node at (-3.6,1.7) {$(b)$};
\end{tikzpicture}
\caption{Concentration thickness $\delta_{c}$ (top panel) for the EN-m configuration and volume of fluid thickness $\delta_\phi$ (bottom panel) for the EN-i configuration. An additional case, NN$_\eta$, is reported for the volume of fluid thickness. Data are plotted up to the stream-wise coordinate where the fluid at the centerline becomes unyielded.}
\label{fig: deltac}
\end{figure}

Figure~\ref{fig: deltac} shows the concentration thickness $\delta_c$ (\reva{panel $(a)$}) and the volume of fluid thickness $\delta_\phi$ (\reva{panel $(b)$}). 
The concentration thickness overlaps with the jet thickness as long as the fluid is yielded (up to $Bi=10^{-2}$), see figure~\ref{fig: qualiuv}. The Schmidt number for the EN-m configuration is $Sc=1$, meaning that the momentum and mass diffusivity are equal; hence we can expect a good overlap of jet and concentration thicknesses in those cases where the jet fluid is fully yielded.
The presence of the unyielded plug increases the concentration thickness, although the increase is lower compared to the increase in the jet thickness. 
A change in the slope of the concentration thickness is observed at the stream-wise coordinate where the unyielded plug begins, see figure~\ref{fig: qualiuv}.
The volume of fluid thickness $\delta_\phi$ shows very little change for increasing Bingham numbers, with a minor increase in $\delta_\phi$ once the plug of unyielded fluid is formed at $Bi=10^{-2}$. Below this threshold Bingham value, the volume of fluid thickness does not show any appreciable change, while above it, the presence of the unyielded plug determines a widening of the jet. The volume of fluid thickness is much lower than the jet and concentration thicknesses due to the absence of diffusion. In addition, the formation of the unyielded plug at high Bingham numbers reduces the spreading rate of the volume of fluid. 
We observe a non uniform trend for the volume of fluid thickness, with the NN configuration showing a larger jet thickness compared to the EN-i $Bi=0$ case; this result is caused by the different viscosity between the EVP jet fluid and the Newtonian ambient fluid. For the NN configuration the viscosity in the jet fluid and in the ambient fluid is uniform and equal to the total viscosity $\eta_t$; for the EN-i configuration the viscosity in the EVP jet fluid is equal to the total viscosity $\eta_t=\eta_p+\eta_s$ whereas the viscosity in the Newtonian ambient fluid is equal to the solvent viscosity $\eta_s$, as there is no contribution from the polymers. 
To verify this hypothesis, an additional Newtonian-Newtonian simulation was performed with non-matched viscosity for the Newtonian jet fluid (viscosity $\eta_t$) and the Newtonian ambient fluid (viscosity $\eta_s$). Data from this additional simulation, labelled as NN$_\eta$, is reported in the bottom panel of figure~\ref{fig: deltac} and in the appendix, figure~\ref{fig: diffvisc}.
It is clear that the case NN$_\eta$ recovers the trend in the volume of fluid thickness; the different jet thickness and centerline velocity observed before have the same origin, as shown in Appendix~\ref{app: diffvisc}. 
Note that, the non-matched viscosity among the EVP jet fluid and the Newtonian ambient fluid affects also the EN-m configuration, but has negligible effects. Indeed, while in the EN-i configuration we have a sharp transition in viscosity at the interface, happening close to the centerline (at a distance $\delta_\phi$) where the shear stresses are large, in the EN-m configuration this transition is smooth, as it changes with the local concentration $\alpha$, and the Newtonian ambient fluid viscosity $\eta_s$ is recovered only further away from the centerline of the jet. This is why the effect of the non-matched viscosity between EVP jet fluid and Newtonian ambient fluid is less important in the EN-m configuration.

\begin{table}[width=.9\linewidth,cols=6,pos=h]
\caption{Power law exponents for the centerline velocity $\beta_u$ and for the jet thickness $\beta_\delta$, and percentage change with respect to the expected exponent, $\Delta\beta_u$ and $\Delta\beta_\delta$. }
\label{tab:expo}
\begin{tabular*}{\tblwidth}{@{} LCCCCC@{} }
\toprule
Configuration & $Bi$ & $\beta_u$ & $\Delta \beta_u$ [\%] & $\beta_\delta$ & $\Delta \beta_\delta$ [\%] \\
\midrule
NN &   &  -0.357  &  7.1  &  0.622  &  -6.7 \\
\midrule
\multirow{5}{*}{EE} & $0$  &  -0.357  &  7.2  &  0.622  &  -6.7 \\
 & $10^{-3}$  &  -0.366  &  9.7  &  0.664  &  -0.4 \\
 & $10^{-2}$  &  -0.419  &  25.8  &  0.729  &  9.4 \\
 & $10^{-1}$  &  -0.996  &  198.8  &  0.843  &  26.4 \\
 & $1$  &  -1.627  &  388.1  &  1.438  &  115.7 \\
\midrule
\multirow{5}{*}{EN-m} & $0$  &  -0.357  &  7.2  &  0.622  &  -6.7 \\
 & $10^{-3}$  &  -0.361  &  8.3  &  0.647  &  -3.0 \\
 & $10^{-2}$  &  -0.374  &  12.1  &  0.640  &  -4.0 \\
 & $10^{-1}$  &  -0.469  &  40.6  &  0.758  &  13.6 \\
 & $1$  &  -0.807  &  142.2  &  0.940  &  41.0 \\
\midrule
\multirow{5}{*}{EN-i} & $0$  &  -0.347  &  4.1  &  0.609  &  -8.6 \\
 & $10^{-3}$  &  -0.353  &  5.9  &  0.637  &  -4.5 \\
 & $10^{-2}$  &  -0.361  &  8.4  &  0.622  &  -6.7 \\
 & $10^{-1}$  &  -0.338  &  1.5  &  0.583  &  -12.5 \\
 & $1$  &  -0.347  &  4.1  &  0.511  &  -23.3 \\
\bottomrule
\end{tabular*}
\end{table}

Finally, we compute the exact power law exponent from our simulation data for the centerline velocity, $\beta_u$, and jet thickness, $\beta_\delta$, obtained with linear regression of $\log u_c = \beta_u\log (x/h) +\log(k)$ for the centerline velocity and $\log \delta_{mom} = \beta_\delta\log (x/h) +\log(k)$ for the jet thickness  (where $k$ are generic offsets) within the stream-wise range where a constant-exponent power law scaling is observed. 
The value of the power law exponents $\beta_u$ and $\beta_\delta$ are reported in table~\ref{tab:expo}, together with the percentage change with respect to the expected exponent for a Newtonian, laminar, planar jet: $\Delta\beta_u=(\beta_u/(-1/3)-1)100$ and $\Delta\beta_\delta=(\beta_\delta/(2/3)-1)100$. 
We confirm that, as long as the jet fluid is yielded, there is limited deviation from the Newtonian power law exponents, while a departure from these is evident once the jet fluid becomes unyielded.



\begin{figure*}[ht!]
\centering
\begin{tikzpicture}
    \node at (0,0) {\includegraphics[width=0.9\textwidth]{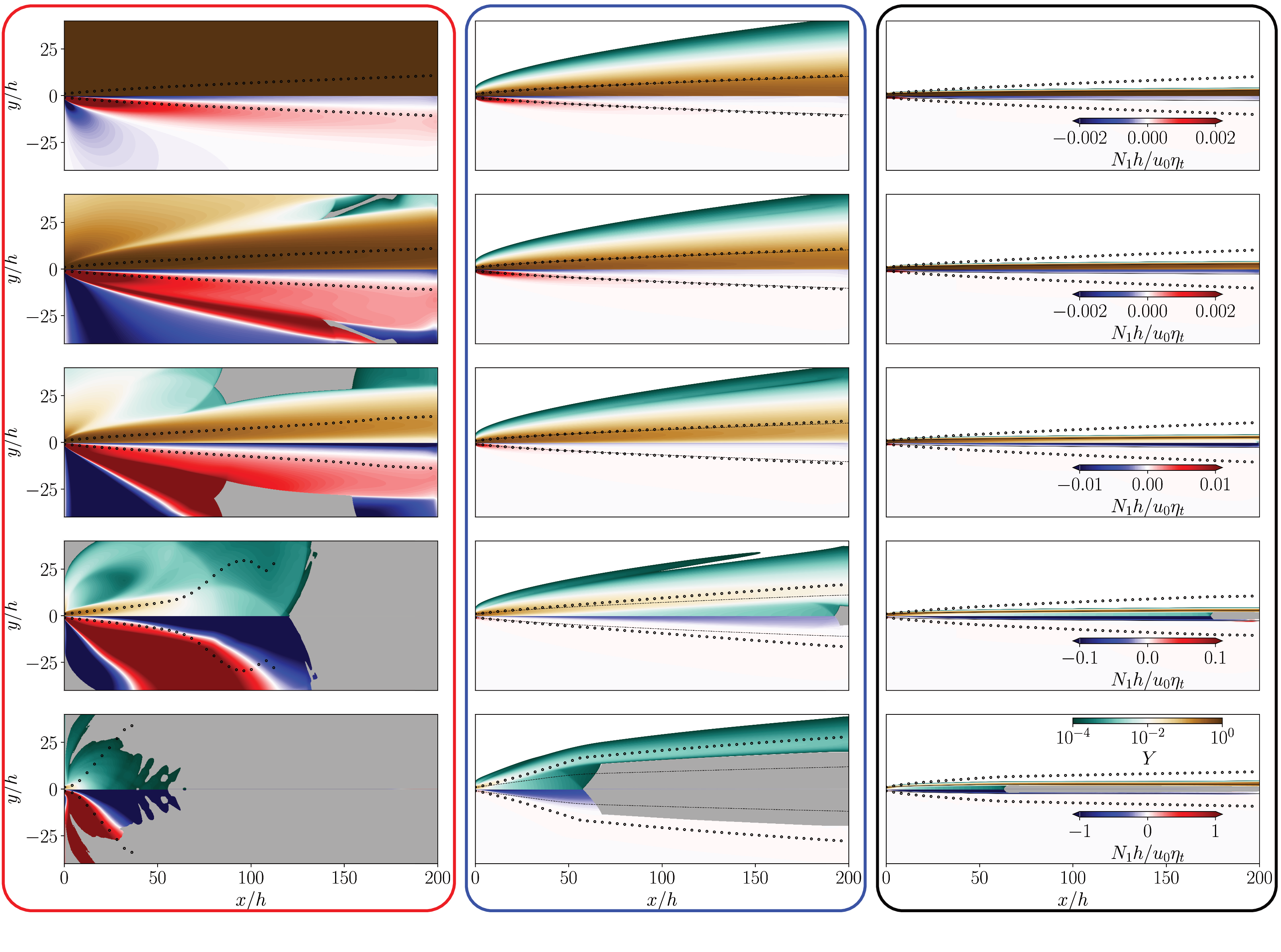}};
    \node at (-6.7,5.3) {$(a)$};
    \node at (-1.7,5.3) {$(f)$};
    \node at (3.3,5.3) {$(k)$};
    \node at (-6.7,3.2) {$(b)$};
    \node at (-1.7,3.2) {$(g)$};
    \node at (3.3,3.2) {$(l)$};
    \node at (-6.7,1.1) {$(c)$};
    \node at (-1.7,1.1) {$(h)$};
    \node at (3.3,1.1) {$(m)$};
    \node at (-6.7,-1) {$(d)$};
    \node at (-1.7,-1) {$(i)$};
    \node at (3.3,-1) {$(n)$};
    \node at (-6.7,-3.1) {$(e)$};
    \node at (-1.7,-3.1) {$(j)$};
    \node at (3.3,-3.1) {$(o)$};
\end{tikzpicture}
\caption{Visualization of the yield variable $Y$ (top half) and normal stress difference $N_1=\tau_{xx}-\tau_{yy}$ (bottom half), with the gray shading identifying unyielded regions. The top and bottom parts of the domain are omitted for clarity. Colored boxes group the different cases: EE (red box), EN-m (blue box) and EN-i (black box). In each box, the Bingham number increases from top ($Bi=0$) to bottom ($Bi=1$). A solid black line identifies the thickness of the volume of fluid, $\delta_\phi$, a dashed black line identifies the thickness of the concentration field, $\delta_c$, and circle black markers identify the jet thickness, $\delta_{mom}$.}
\label{fig: qualiyield}
\end{figure*}

Next, we turn our attention to the non-Newtonian stress tensor. We report in figure~\ref{fig: qualiyield} the yield variable $Y$ (top half) and the first normal stress difference $N_1=\tau_{xx}-\tau_{yy}$ (bottom half). 
For the EE configuration, we observe a reduction in the value of the yield variable for increasing Bingham numbers throughout the domain. At $Bi=0$ the yield variable is $Y=1$ everywhere, as expected; as the Bingham number is increased, regions characterized by low $Y$ values and unyielded fluid form outside the jet, and for $Bi\ge 10^{-1}$, the region of unyielded fluid reaches the jet centerline and further extends towards the inlet. 
Also, the yield variable is largest near the inlet and in the shear layer of the jet.
In the EN-m configuration, we observe very little change in the $Y$ field at low values of the Bingham number, as long as the fluid is yielded: the maximum values are achieved within the shear layer of the jet and $Y$ smoothly decays away from the centerline of the jet. Once an unyielded plug is formed in the core of the jet, a region characterized by very low $Y$ values forms in the core of the jet and approaches zero in the core of the jet. 
This behaviour has two competing contributions: a reduction in the values of the stresses in the core and in the far field of the jet, which leads to a reduction in $Y$, and a reduction in the concentration moving away from the centerline of the jet, which causes an increase in $Y$ due to a reduction in the yield stress. Thus, in the core of the jet where the concentration is largest, the yield stress overcomes the deviatoric stress $|\tau_d|$, which is relatively small, hence the formation of an unyielded plug is favorable. In the far field of the jet, the deviatoric stress is small again, but the yield stress is reduced due to the low concentration of the EVP fluid; results show that the yield stress value decays faster than the deviatoric stress, as no unyielded region appears away from the jet centerline. In the shear layer of the jet (at $Bi=1$ this region corresponds to the fluid surrounding the unyielded plug), the increase in the deviatoric stress outweighs the reduction in the yield stress, as shown by the local maximum of $Y$ along the stream-wise coordinate. 
The immiscible EVP-Newtonian configuration differs in this aspect as the local concentration is uniform and unitary within the jet and zero otherwise: there is no smooth decay along the jet-normal direction nor the stream-wise direction. At $Bi=0$ the yield variable is equal to one within the jet, with a fast transition at the interface between the EVP and the Newtonian fluid. As the Bingham number is increased, a decrease in the value of the yield variable is reported in the core of the jet (starting from the outlet and extending towards the inlet for increasing $Bi$); the deviatoric stress is lowest in the core of the jet and reduces along the stream-wise direction. As the Bingham number is increased, the yield stress increases, thus reducing the value of the yield variable and leading to the formation of an unyielded plug at $Bi=10^{-2}$. A region of yielded fluid encapsulates the unyielded plug, due to the large deviatoric stress at the jet boundary (at about $y=\pm\delta_\phi$). 

\begin{figure*}[ht!]
\centering
\begin{tikzpicture}
    \node at (0,0) {\includegraphics[width=0.9\textwidth]{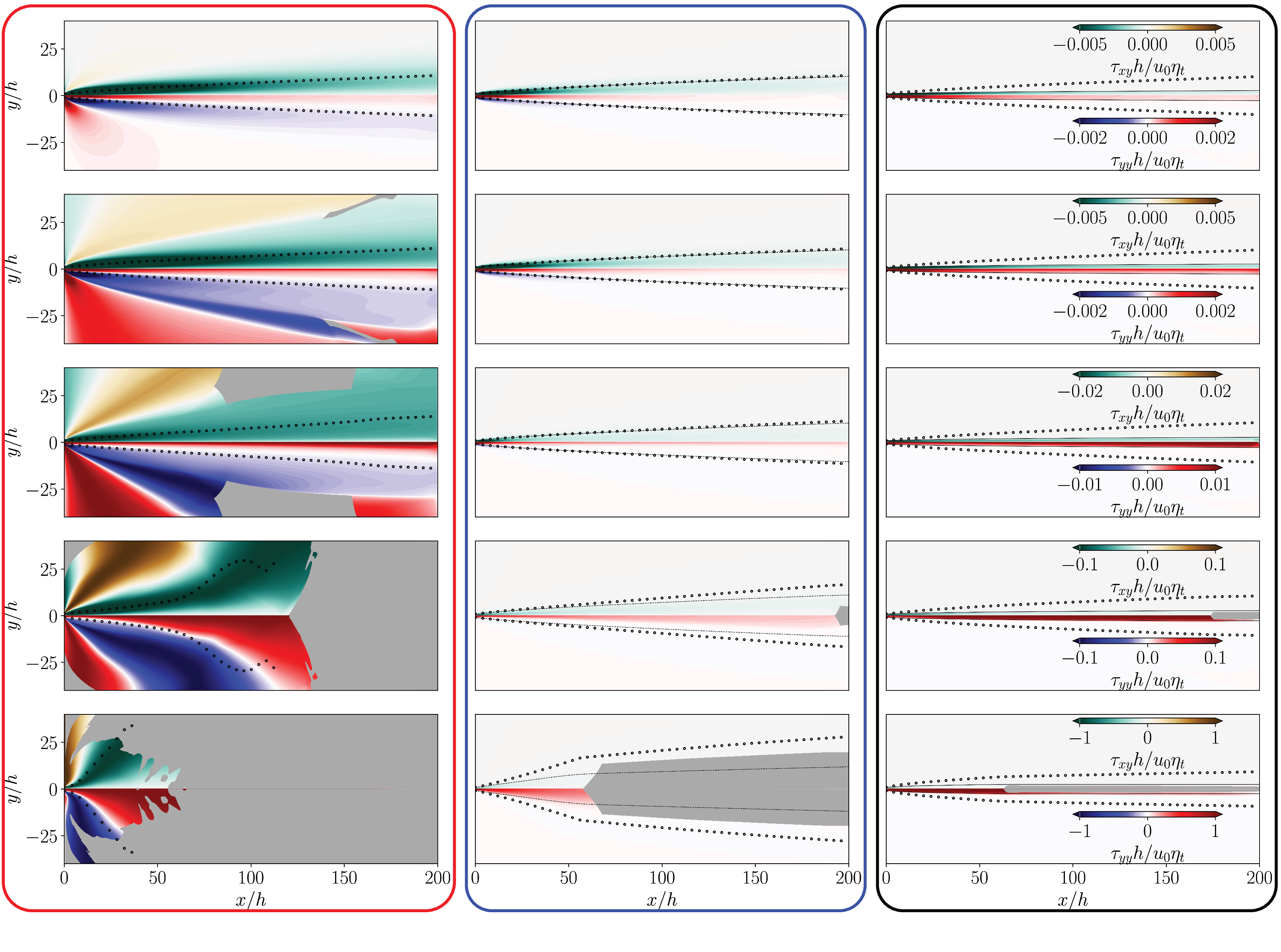}};
    \node at (-6.7,5.3) {$(a)$};
    \node at (-1.7,5.3) {$(f)$};
    \node at (3.3,5.3) {$(k)$};
    \node at (-6.7,3.2) {$(b)$};
    \node at (-1.7,3.2) {$(g)$};
    \node at (3.3,3.2) {$(l)$};
    \node at (-6.7,1.1) {$(c)$};
    \node at (-1.7,1.1) {$(h)$};
    \node at (3.3,1.1) {$(m)$};
    \node at (-6.7,-1) {$(d)$};
    \node at (-1.7,-1) {$(i)$};
    \node at (3.3,-1) {$(n)$};
    \node at (-6.7,-3.1) {$(e)$};
    \node at (-1.7,-3.1) {$(j)$};
    \node at (3.3,-3.1) {$(o)$};
\end{tikzpicture}
\caption{Non-Newtonian extra-stresses: $\tau_{xy}$ (top half) and $\tau_{yy}$ (bottom half). Gray shading identifies unyielded regions. Colored boxes group the different cases: EE (red box), EN-m (blue box) and EN-i (black box). In each box, the Bingham number increases from top ($Bi=0$) to bottom ($Bi=1$). A solid black line identifies the thickness of the volume of fluid, $\delta_\phi$, a dashed black line identifies the thickness of the concentration field, $\delta_c$, and circle black markers identify the jet thickness, $\delta_{mom}$.}
\label{fig: qualies}
\end{figure*}

\revb{In the present configuration, the two normal stress differences are not null, due to the finite, albeit small, value of elasticity considered.} The first normal stress difference $N_1=\tau_{xx}-\tau_{yy}$ is reported in the bottom half of figure~\ref{fig: qualiyield}; figure~\ref{fig: qualies} shows the non-Newtonian extra-stresses, the off-diagonal component $\tau_{xy}$ (top half) and the jet-normal component $\tau_{yy}$ (bottom half). 
Note that, the normal stresses $\tau_{xx}$ (stream-wise, not shown here) and $\tau_{yy}$ (jet-normal, figure~\ref{fig: qualies} -- bottom half) have a very similar distribution albeit with opposite sign, \textit{i.e.} $\tau_{xx} \simeq -\tau_{yy}$, thus resulting in the first normal stress difference being $N_1 \simeq 2 \tau_{xx}$. For this reason, we report only the first normal stress difference, the jet-normal and off-diagonal components of the extra-stress tensor. 
\revb{Also, note that the magnitude of the extra-stresses increases with the Bingham number for all configurations, due to the reducing values of the yield variable. 
To better understand the role of the yield variable $Y$, we can rewrite the constitutive equation of the EVP fluid, equation~\ref{eq:nntau}, for the yielded fluid case (\textit{i.e.}, $Y>0$) as:
\begin{equation}
\lambda^* \overset{\nabla}{\boldsymbol{\tau}}+\boldsymbol{\tau}=\eta_p^*\left( \nabla\mathbf{u} +\nabla\mathbf{u}^T \right).
\label{eq:oldb}
\end{equation}
The parameters $\lambda^*$ and $\eta_p^*$ are the relaxation time scale and the polymer viscosity re-scaled by the yield variable: $\lambda^*=\lambda/Y \ge \lambda$ and $\eta_p^*=\eta_p/Y \ge \eta_p$. Equation~\ref{eq:oldb} can be understood as the constitutive equation for an Oldroyd-B fluid with relaxation time $\lambda^*$ and polymer viscosity $\eta_p^*$, in which the yield variable has changed the model parameters of the fluid. 
This increase in the relaxation time and polymer viscosity results then in a increase in magnitude of the non-Newtonian extra-stresses for increasing Bingham values.}

In the EE configuration, the stream-wise extra-stress is negative at the jet centerline (negative $N_1$) and the jet-normal extra-stress is positive at low Bingham numbers. This region massively extends away from the jet centerline as the fluid within the jet becomes unyielded, for $Bi\ge10^{-1}$. As we move away from the centerline of the jet, the stream-wise component changes sign and becomes negative and once again positive in the ambient fluid; the jet-normal component has a similar distribution although with opposite sign. Interestingly, we observe that the normal components change sign across any region of unyielded fluid  at intermediate values of the Bingham number, see for example the cases at $Bi=10^{-3}$ and $Bi=10^{-2}$. 
The off-diagonal shear component $\tau_{xy}$ (top half of the domain) is zero at the centerline due to the symmetry of the flow, negative within the jet and positive (with small value) in the ambient fluid surrounding the jet flow. 
A similar distribution of the non-Newtonian extra-stresses is reported for the EN-m configuration: the core of the jet is characterized by a negative stream-wise extra-stress ($\tau_{xx}$) and a positive jet-normal extra-stress ($\tau_{yy}$). The extra-stresses change sign at a distance from the centerline of about one concentration thickness $\delta_c$; this effect is limited to the low Bingham numbers and disappears at larger values of the Bingham number. 
The off-diagonal shear component of the extra-stress tensor, $\tau_{xy}$, is negative at distances from the centerline smaller than $\delta_c$. 
Non-Newtonian extra-stresses are strictly limited to the jet fluid only in the EN-i configuration, as the EVP and Newtonian fluids cannot mix. 
The stream-wise component of the extra-stress tensor is negative everywhere but for a small region near the inlet, where it becomes positive at small Bingham numbers. The same distribution, with opposite sign, is reported for the jet-normal component of the extra-stresses. The shear component of the extra-stress tensor, $\tau_{xy}$, is negative everywhere and achieves its maximum value at distances from the centerline slightly smaller than the volume of fluid thickness $\delta_\phi$. 
Overall, the EE configuration has the highest magnitude of the extra-stresses across the different configurations, while the extra-stresses in the EN-m configuration are the smallest.

\begin{figure}[ht!]
\centering
\begin{tikzpicture}
    \node at (0,14.1) {\includegraphics[width=0.9\columnwidth]{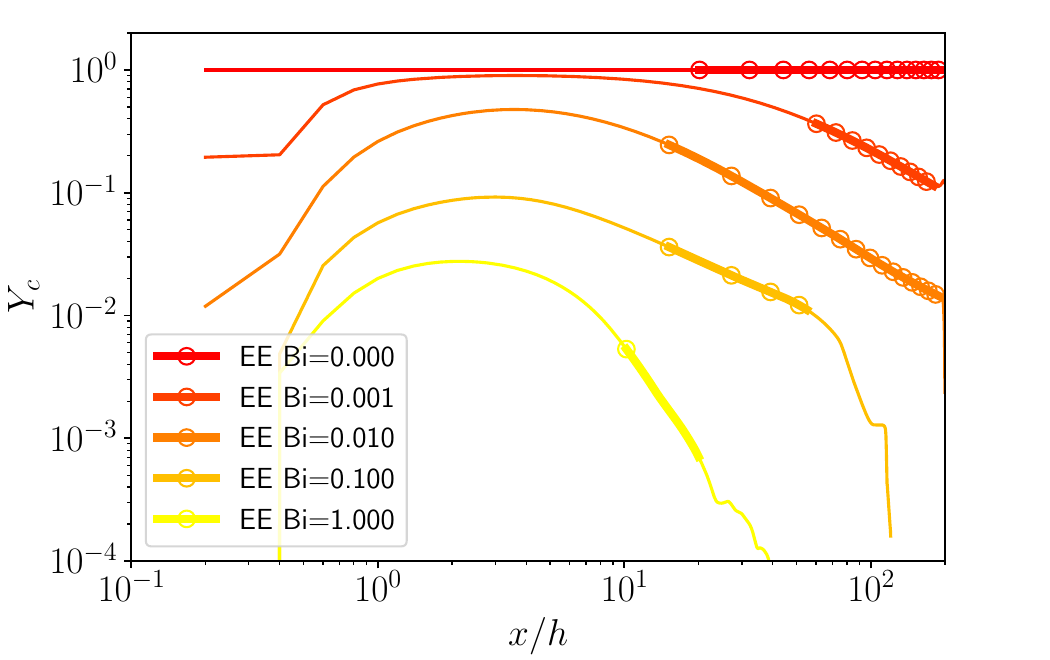}};
    \node at (0,9.4) {\includegraphics[width=0.9\columnwidth]{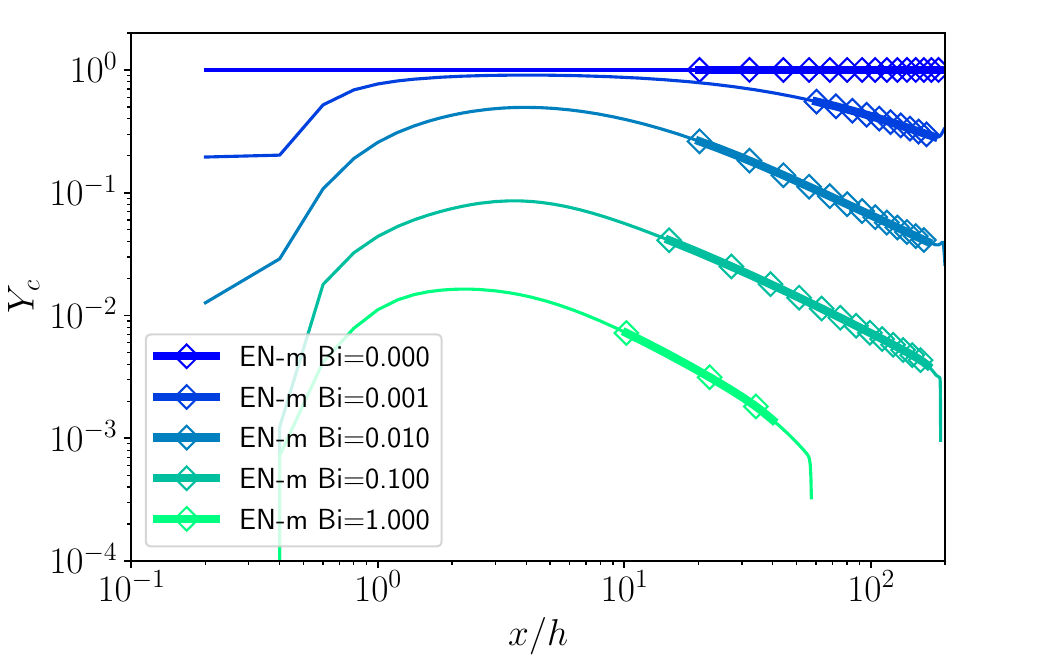}};
    \node at (0,4.7) {\includegraphics[width=0.9\columnwidth]{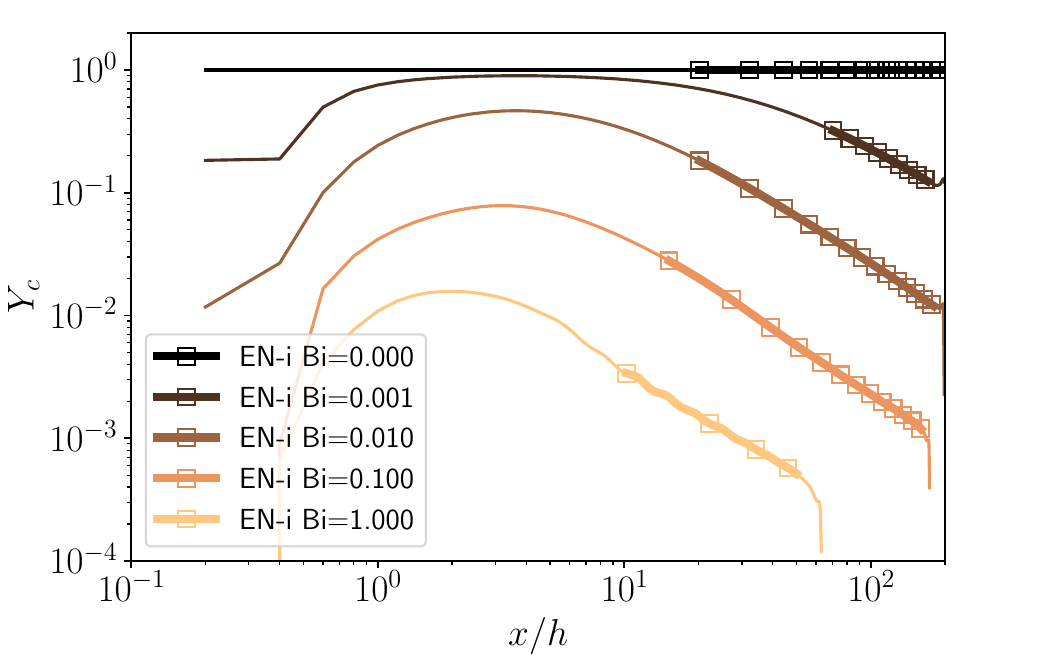}};
    \node at (0,0.0) {\includegraphics[width=0.9\columnwidth]{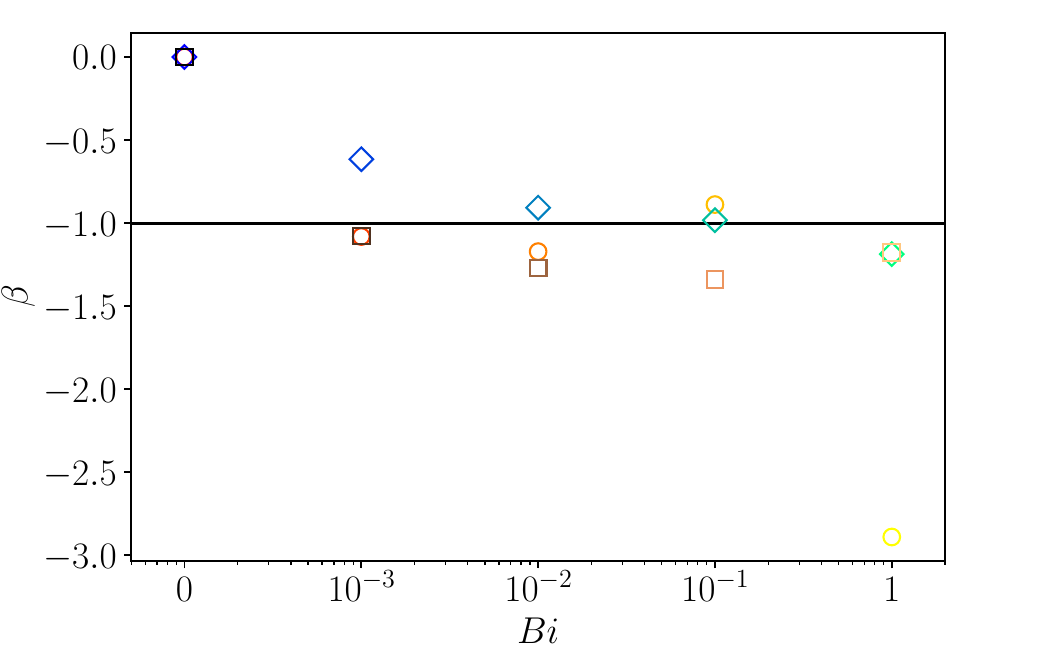}};
    \node at (-3.6,16.) {$(a)$};
    \node at (-3.6,11.3) {$(b)$};
    \node at (-3.6,6.4) {$(c)$};
    \node at (-3.6,1.7) {$(d)$};
\end{tikzpicture}
\caption{Yield variable at the centerline of the jet, $Y_c$. Each panel refers to a different configuration (top to bottom: EE, EN-m and EN-i). Data are plotted along the stream-wise direction as long as the fluid at the centerline is yielded. A thicker line with markers identifies the region characterized by power law scaling, where linear regression has been performed. In the last panel, the power law exponent $\beta$ (defined as $Y_c\propto (x/h)^\beta$) is reported as a function of the Bingham number.}
\label{fig: yield}
\end{figure}
Before concluding, we show the yield variable at the centerline $Y_c$ in figure~\ref{fig: yield}, where each panel corresponds to a different configuration (EE, EN-m, EN-i) and the last panel reports the power law exponent $\beta$.
For all cases with $Bi>0$, the yield variable at the centerline is a convex function that has low values at the inlet, as the EVP jet fluid is injected with zero extra-stresses at the inlet boundary, and progressively grows in the stream-wise direction as the jet flow evolves. The centerline yield variables achieves its maximum value relatively close to the inlet, at a stream-wise distance of about $3$ to $5h$, and smoothly decays further downstream. 
The position of the maximum shifts towards the inlet for increasing values of the Bingham number, with the different configuration (EE, EN-m or EN-i) having negligible effect on the position and value of the maximum. 
When reported in logarithmic scale, the presence of a power law decay in the fully-developed region of the jet ($x>10h$) is clear. 
We assume the centerline yield variable in the fully-developed region to scale as $Y_c\propto (x/h)^{\beta}$, and use linear regression on $\log Y_c = \beta\log (x/h) +\log(k)$ (the centerline yield variable is approximated as $Y_c=k(x/h)^{\beta}$, with $k$ being a scale factor) to compute the power law exponent $\beta$.
Linear regression is performed over a stream-wise range where a linear power-law scaling is observed, indicated by a thicker line width with markers in figure~\ref{fig: yield}.
Excluding the trivial cases at $Bi=0$ and the case EE at $Bi=1$ (when almost all fluid is unyielded), the power law exponent is $\beta \approx 1$. While for the EVP-Newtonian cases (both miscible and immiscible) there seems to be a trend with the Bingham number (with $\beta$ decreasing with growing Bingham numbers), the computed values of the power law exponent for the EE cases fluctuate around minus one.
The observed scaling in the yield variable at the centerline, can be derived as follows. With the hypothesis of yielded fluid, for which $|\tau_d|>\tau_y$, we decompose the magnitude of the deviatoric stress as $|\tau_d|=\tau_y +\Delta\tau$, with $\Delta \tau$ being positive. Thus, we can rewrite the centerline yield variable $Y_c$ for $|\tau_d|>\tau_y$ as:
\begin{equation}
Y_c=\frac{|\tau_d| -\tau_y}{|\tau_d|}=\frac{\Delta\tau}{\tau_y+\Delta\tau}=\left(1+\frac{\tau_y}{\Delta\tau}\right)^{-1}.
\end{equation}
The stress $\Delta \tau$ is expected to scale with the local variables (centerline velocity and jet thickness) as $\Delta\tau\propto \eta_t u_c/\delta_{mom}$, and, by introducing the scaling found for $u_c$ and $\delta_{mom}$, we obtain $\Delta\tau\propto (x/h)^{-1}$. Reworking this result into the yield variable at the centerline, we obtain $Y_c \propto (x/h)^{-1}$, whose exponent $-1$ is consistent with those found from our numerical data.

\section{Conclusions}
\label{sec: concl}
We investigate the dynamics of non-Newtonian planar jets at low Reynolds number via numerical simulations. The non-Newtonian fluid is modeled by the \citet{saramito2007new} elastoviscoplastic fluid model. 
We consider three different configurations: an EVP jet in EVP ambient fluid, an EVP jet in Newtonian ambient fluid (miscible fluids), and an EVP jet in Newtonian ambient fluid (immiscible fluids).

As the Bingham number is increased, larger fractions of the fluid become unyielded, thus affecting the overall dynamics of the jet.
When the jet fluid is yielded (\textit{i.e.}, at low values of the Bingham number), the non-Newtonian jet shows very similar dynamics to the Newtonian jet (NN) for all the three configurations considered here. The centerline velocity and jet thickness overlap with the statistics from the NN configuration, and the the power law scalings obtained for laminar, Newtonian, planar jets \citep{Mei_2001,soligo2023non} are found. 
The formation of regions of unyielded fluid for increasing values of the Bingham number leads to a deviation from the NN configuration: an increase in the spreading rate of the jet and in the decay rate of the centerline velocity is observed, induced by regions of unyielded fluid at the outlet trough a blockage effect. While spreading rate and centerline velocity decay rate are increased, the same power law exponents of the NN configuration are retained.
The highest jet spreading rate and centerline velocity decay rate are observed for the EE configuration, followed by the EN-m configuration and lastly by the EN-i configuration. The extent of the unyielded fluid regions determines this trend: for the EE configuration both the jet and ambient fluid are unyielded, thus causing the strongest blockage effect among the three configurations, whereas for the EN configurations only the jet fluid (and surrounding, for the miscible case) are unyielded.
We report an increase of the non-Newtonian extra-stress for increasing values of the Bingham number; 
the normal components of the non-Newtonian extra-stress tensor ($\tau_{xx}$ and $\tau_{yy}$) are equal in magnitude but opposite in sign, leading to the first normal stress difference to be $N_1\simeq2\tau_{xx}$.
The largest non-Newtonian extra-stresses are reported for the EE configuration, whereas the EN-m configuration is characterized by the lowest extra-stresses across the three configurations for the same value of the Bingham number.
A power law scaling for the yield variable at the centerline is derived, $Y_c\propto (x/h)^{-1}$, with a fair agreement between the derived scaling law and the simulation results for all the configurations.

\section*{Acknowledgements}
The research was supported by the Okinawa Institute of Science and Technology Graduate University (OIST) with subsidy funding from the Cabinet Office, Government of Japan. The authors acknowledge the computer time provided by the Scientific Computing and Data Analysis section of the Core Facilities at OIST, and the computational resources on SQUID provided by the Cybermedia Center at Osaka University through the HPCI System Research Project (project ID: hp230018).

\appendix

\section{Effect of non-matched viscosity}
\label{app: diffvisc}

\begin{figure}[ht!]
\centering
\includegraphics[width=0.9\columnwidth]{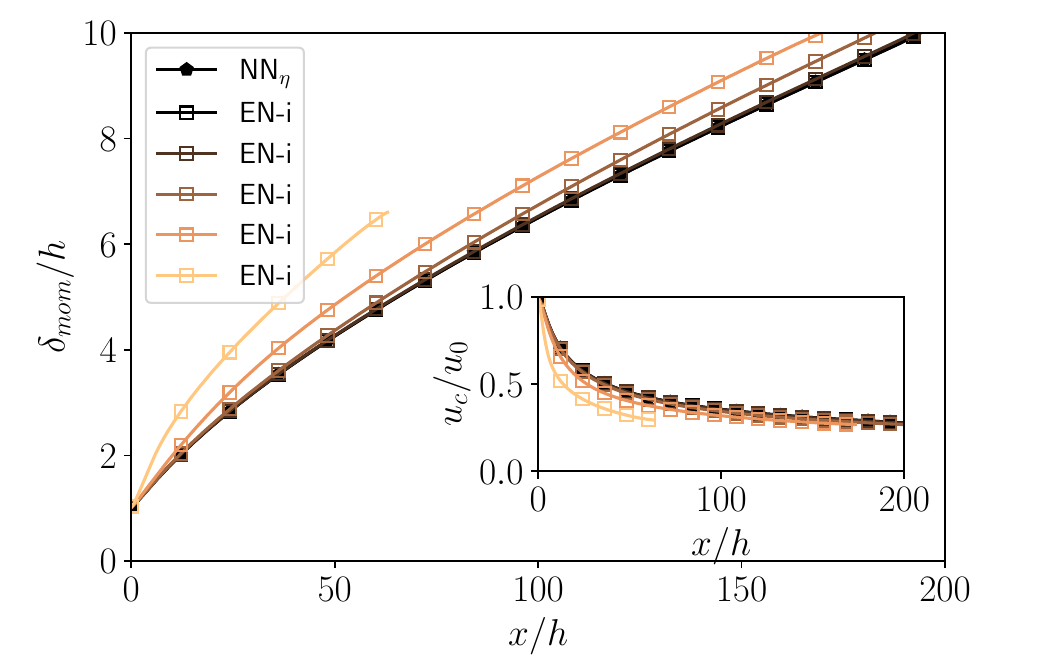}
\caption{Jet thickness $\delta_{mom}$ and centerline velocity $u_c$ (inset) for the EN-i configuration, together with the additional configuration NN$_\eta$. Data are plotted up to the stream-wise coordinate where the fluid at the centerline becomes unyielded.}
\label{fig: diffvisc}
\end{figure}

We report here the jet thickness and centerline velocity for the EN-i configuration together with the data from the NN$_\eta$ case. The NN$_\eta$ case is a Newtonian jet fluid issued into Newtonian ambient fluid, with non-matched viscosity among the jet fluid (viscosity $\eta_t$) and the ambient fluid (viscosity $\eta_s$). We demonstrate that the discrepancy observed in the centerline velocity, figure~\ref{fig: vc}, and in the jet thickness, figure~\ref{fig: delta}, is caused by the difference in viscosity in the EN-i configuration between the jet fluid and the ambient fluid. For the NN configuration, both the jet fluid and the ambient fluid have the same viscosity, $\eta_t$, whereas the NN$_\eta$ configuration is characterized by different viscosities for the jet and the ambient fluid. The NN$_\eta$ configuration is thus more representative of a reference case for the EN-i configuration, where the jet fluid (EVP, viscosity $\eta_t$) and the ambient fluid (Newtonian, viscosity $\eta_s$) have different viscosity. 
When comparing the NN$_\eta$ and the EN-i at low Bingham number, a collapse of the data for the jet thickness and centerline velocity is observed in figure~\ref{fig: diffvisc}.

\begin{figure}[ht!]
\centering
\begin{tikzpicture}
    \node at (0,4.8) {\includegraphics[width=0.9\columnwidth]{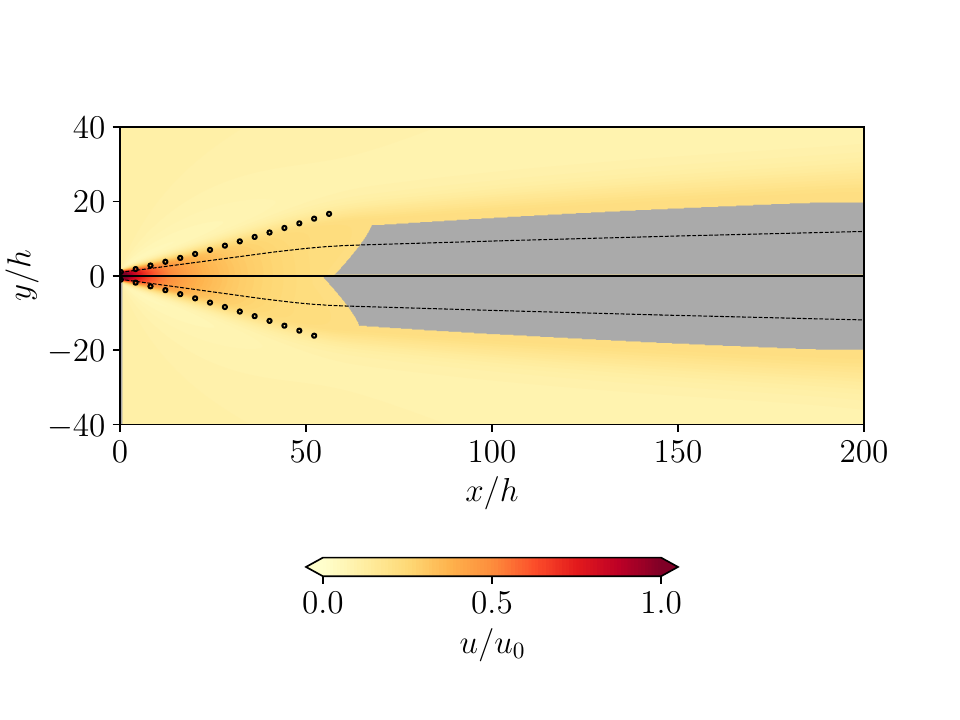}};
    \node at (0,0.0) {\includegraphics[width=0.9\columnwidth]{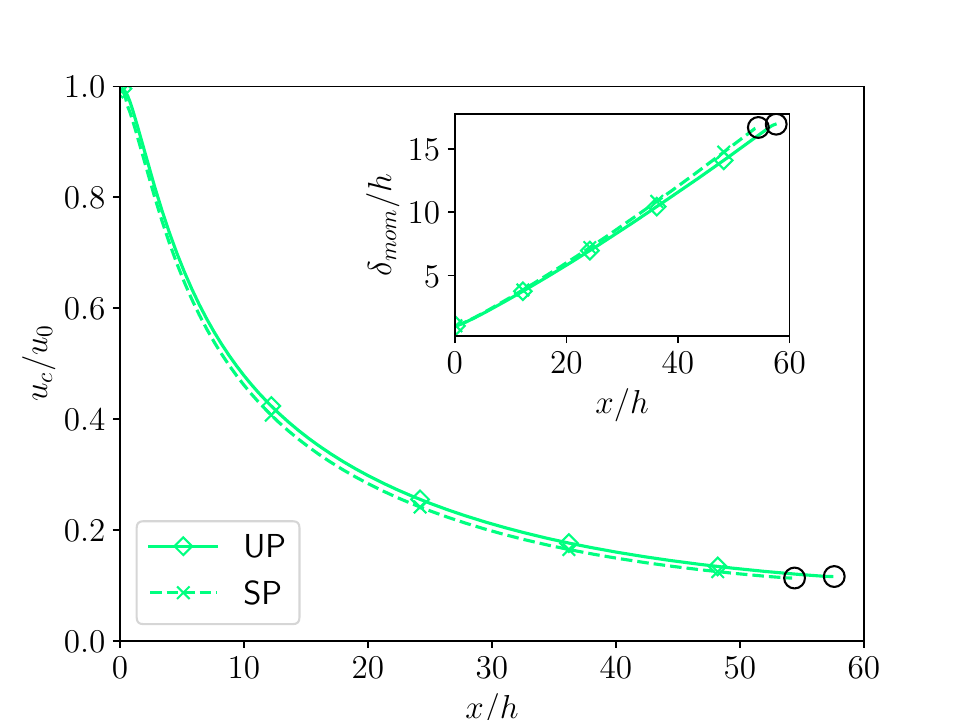}};
    \node at (-3.6,6.4) {$(a)$};
    \node at (-3.6,1.8) {$(b)$};
\end{tikzpicture}
\caption{Panel $(a)$: comparison of stream-wise velocity for the case with zero extra-stress at the inlet (label UP, top half of the domain $y/h>0$) and for the case with non-zero extra-stresses at the inlet (label SP, bottom half of the domain $y/h<0$). A thin dashed line identifies the concentration thickness $\delta_c$ and black circles identify the momentum thickness $\delta_{mom}$. Panel $(b)$: comparison of centerline velocity and jet thickness (inset). Black circle markers identify the coordinate where the fluid becomes unyielded.}
\label{fig:compare}
\end{figure}

\revb{
\section{Effect of non-zero polymer extra-stresses at the inlet}
\label{app:inlet}
We tested the effect of non-zero polymeric extra-stresses at the inlet on the overall jet dynamics. An additional simulation with the same parameters of the case EN-m at $Bi=1.0$ has been performed. In this additional simulation a different inlet boundary condition has been considered for the polymer extra-stresses $\boldsymbol{\tau}$: the extra stresses at the inlet were computed based on the imposed inlet velocity profile. The reference case ($\boldsymbol{\tau}=\mathbf{0}$ at the inlet, labelled as unstretched polymers UP) and the new case with polymer extra-stresses at the inlet (labeled as stretched polymers SP) are reported in figure~\ref{fig:compare}. The stretched polymers case shows a slightly earlier transition to unyielded fluid compared to the unstretched polymer case, see the comparison in figure~\ref{fig:compare}$(a)$; the shape and thickness of the unyielded fluid is however unaffected by the different conditions on the polymeric stresses at the inlet. The bulk statistics of the jet reported in figure~\ref{fig:compare}$(b)$, i.e., the centerline velocity and jet thickness, show overall minimal differences among the two cases. We can thus confirm that different values of the polymer extra-stresses at the inlet have minimal effects on the overall dynamics of the jet and on the yielded and unyielded regions. }

\bibliographystyle{cas-model2-names}

\bibliography{jnnfm}

\end{document}